\documentclass[referee]{aa}
\usepackage[dvips]{graphicx}

\begin{document}

\title{Time-resolved spectroscopy and photometry of the dwarf nova FS Aurigae
in quiescence
\thanks{Based on observations made at the Special Astrophysical
Observatory, Nizhnij Arkhyz, Russia.}}

\author{V.V. Neustroev}


\institute{
           Department of Astronomy and Mechanics, Udmurtia State University,
           Universitetskaia, 1, Izhevsk, 426034, Russia \\
           e-mail: benj@uni.udm.ru
}

\date{Received , accepted }

\titlerunning{Time-resolved spectroscopy and photometry of FS Aur}
\authorrunning{V.V. Neustroev}

\abstract{
We present results of non-simultaneous time-resolved photometric and spectroscopic 
observations of the little-studied dwarf nova FS Aur in quiescence. 
The spectrum of FS Aur shows strong and broad emission lines of hydrogen and 
He\,I, and of weaker He\,II $\lambda4686$ and C\,III/N\,III blend, similar to 
other quiescent dwarf novae. All emission lines are single-peaked, however their 
form varies with orbital phase. Absorption lines from a late-type secondary are 
not detected. From the radial velocity measurements of the hydrogen lines H$_\beta$ and 
H$_\gamma$ we determined a most probable orbital period P=0\fd059 $\pm$ 0\fd002. 
This period agrees well with the 0\fd0595 $\pm$ 0\fd0001 
estimate by Thorstensen et al. (1996). On the other hand, the period of 
photometric modulations is longer than the spectroscopic period and can be 
estimated as 3 hours. Longer time coverage during a single night is 
needed to resolve this problem.
Using the semi-amplitude of the radial velocities, obtained from measurements
of hydrogen and helium lines, and some empirical and theoretical relations 
we limited the basic parameters of the system: a mass ratio $q \geq 0.22$, a primary mass 
$M_{1}=0.34 - 0.46 M_{\sun}$, a secondary mass $M_{2} \leq 0.1M_{\sun}$,
and an inclination angle $i=51^{\circ } - 65^{\circ}$.
Doppler tomography has shown at least two bright regions in the accretion disk of FS Aur.
The first, brighter spot is located at phase about 0.6. The second spot is located 
opposite the first one and occupies an extensive area 
at phases about $0.85 - 1.15$. 
      \keywords{accretion, accretion disks --
                stars: binaries: spectroscopic --
                stars: cataclysmic variables --
                stars: individual: FS Aur
               }
}

\maketitle


\section{Introduction}

FS Aur was discovered and first classified as a dwarf nova by Hoffmeister (\cite{hoff}).
This little-studied system varies between aproximately V=15$\fm$4--16$\fm$2 in
quiescence and V=14$\fm$4 in outburst. Optical photometry during quiescence was
reported by Howell \& Szkody (\cite{how:szkody}) who found a modulation with period
P=97$\pm$10 minutes, characteristic for SU Uma - type stars.
The long
term light curve was analysed by Andronov (\cite{andronov}). He found normal outbursts
only with a probable interval between them of 12$^d$ and no superoutburst.
Therefore the SU Uma classification of this star is tentative.

Spectroscopic observations by Williams (\cite{williams}) showed a typical dwarf nova
spectrum with strong emission lines of hydrogen and He\,I. H$_\alpha $
velocity variations with a period of 85.7$\pm$0.18 min have been reported by
Thorstensen et al. (\cite{TPST}; TPST hereafter). No other study of emission lines
has been made
for FS Aur. This motivated us to perform time-resolved spectroscopy and
photometry of FS Aur in order to study its properties in more detail.

\begin{figure}
\resizebox{\hsize}{!}{\includegraphics{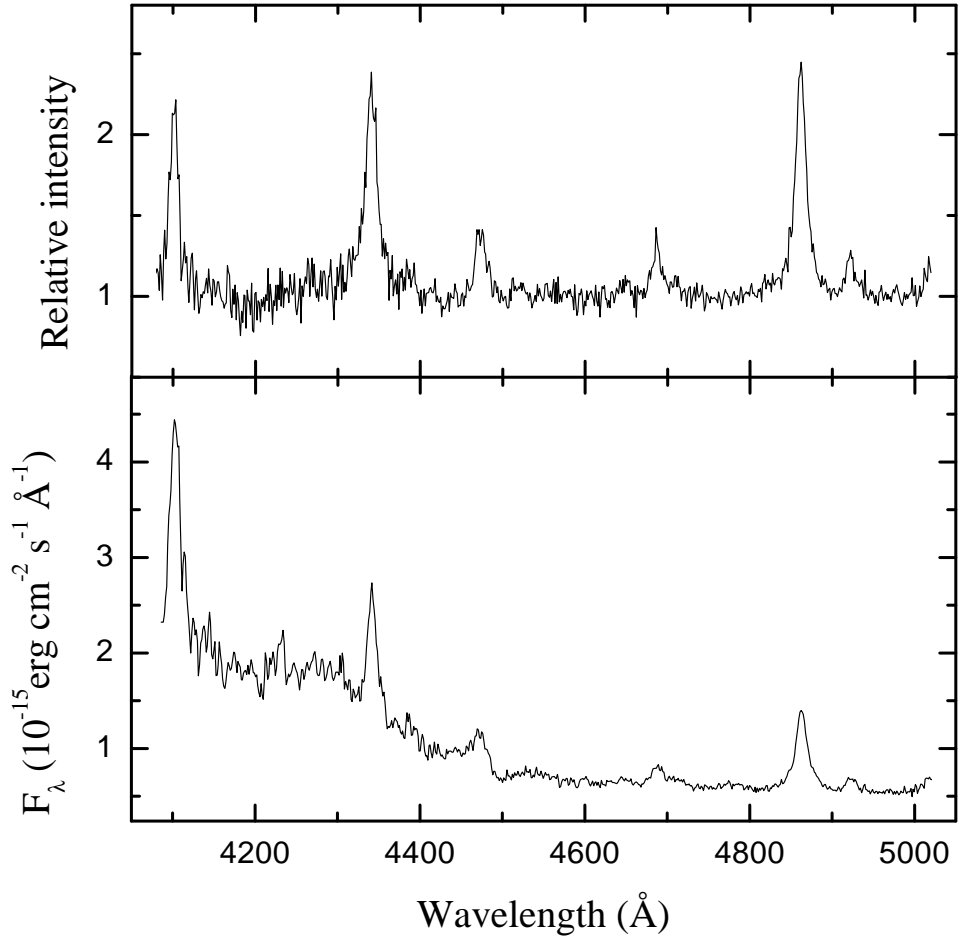}}
\caption{The single typical normalized spectrum (top panel) and the 
average spectrum (bottom panel) of FS Aur. The mean spectrum is an average 
of all spectra, corrected for wavelength shifts due to orbital motion.}
\label{aver_spec}
\end{figure}


\section{Observations and data reduction}

\subsection{Spectroscopy}

The spectroscopic observations of FS Aur were obtained on January 18,
1996 with the 6 meter telescope at the Special Astrophysical Observatory.
The SP-124 spectrograph was used with a
PHOTOMETRICS CCD, which has 1024$\times $1024 pixels. The seeing was 
around 2 arcseconds and we selected a slit width of 1$\farcs$5. The
spectra have a resolution of 2.6 \AA, covering the range 4100-5000~\AA .
About 1.5 orbital cycles were covered with 29 spectra of 200 sec long exposure
(dead time between exposures was 30 seconds). He-Ne-Ar lamp exposures were
taken typically every 30 min.

The spectra were reduced in the standard manner. All CCD frames were
debiased, flat-fielded and wavelength calibrated using the MIDAS system. For
the wavelength calibration of the spectra, interpolation was used between
neighbouring arc spectra. The root mean square of the polynomial fits is $%
\sim $0.030 \AA . One-dimensional spectra were extracted using the optimized
algorithm proposed by Horne (\cite{horne}). The resulting spectra were reduced to an
absolute flux scale by calibration with standard star G191b2, which was
observed in the same night.
Note that substantial light losses at the slit edges occur.
Therefore, our data cannot be used to obtain the absolute fluxes though 
the obtained fluxes are good to approximately 10\%. In view of this, from 
now on we shall use spectra normalized to the continuum.


\subsection{Photometry}

The photometric observations were obtained with the 1 meter telescope at the
Special Astrophysical Observatory (Nizhnij Arkhyz, Russia), using a
1024$\times $1024 CCD camera. 31
exposures in the V filter with 180 seconds integration time were obtained on
February 12, 1997. The total observation time was more than 2 hours.
Additional V-band CCD observations were carried out on October
11, 1997 during 4 hours using the same telescope with a 530$\times $580 CCD
camera. A total of 34 frames were recorded with an integration time of 300 sec.

All CCD images were debiased and flat-fielded using the MIDAS system. We
calculated magnitudes of the source with respect to several ``comparison''
stars within the field of view from Misselt (\cite{misselt}), using aperture
photometry (we used standard stars C$_{1}$, C$_{3}$ and C$_{4}$).
The differences between daily averages of the differential magnitudes
(C$_{1}$--C$_{3}$, C$_{4}$--C$_{1}$, C$_{4}$--C$_{3}$)
were within 0.01 mag during our observational runs.


\begin{figure}[t]
\vspace{0.5 cm}
\resizebox{\hsize}{!}{\includegraphics{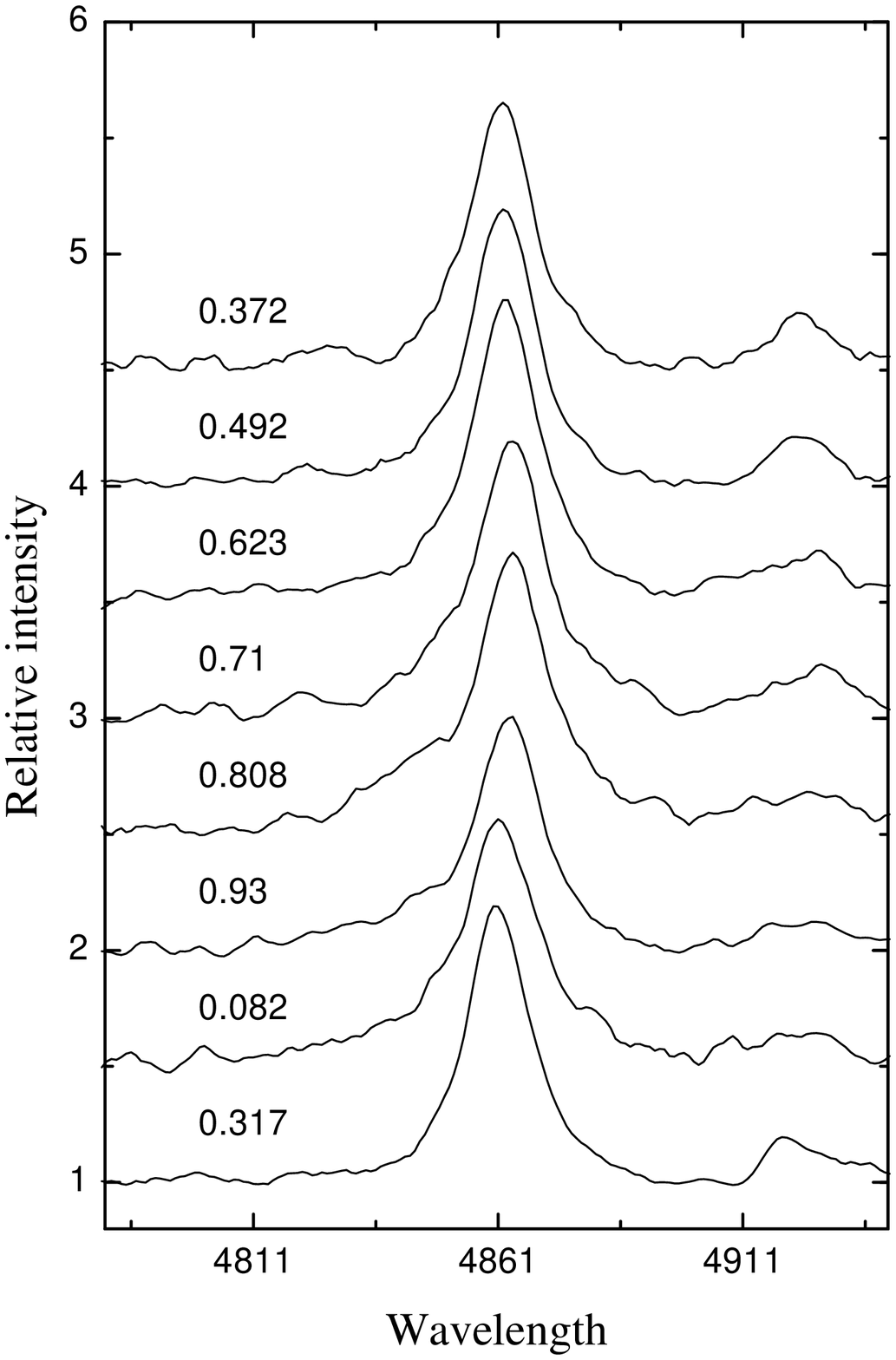}}
\caption{The variation of the H$_{\beta }$ line profile around the orbit.
The presence of a broad base component and a narrow line component is evident. 
There are orbital variations in the line profile. The narrow component remains 
practically symmetrical throughout an orbital cycle, but on the blue wing of 
the broad component at phases about 0.75 -- 0.9 there appears a hump.
}
\label{profiles}
\end{figure}

\section{Data analysis and results}


\subsection{The average spectrum}

The single typical spectrum and the average spectrum of FS Aur are shown in
Fig.~\ref{aver_spec}. The mean spectrum is an average of all spectra, corrected for
wavelength shifts due to orbital motion. The spectrum is typical for a dwarf
nova. It is dominated by strong and broad emission lines of
hydrogen and neutral helium.
In addition to them He\,II $\lambda4686$ and weak C\,III/N\,III blend are  
observed also.
All emission lines are single-peaked. The
Balmer decrement is flat, indicating that the emission is optically thick as
is normal in dwarf novae.
There is no evidence of a contribution from a late-type secondary star.
The equivalent width, FWHM, FWZI, and Relative
Intensity of the major emission lines are presented in Table ~\ref{Table1}.

\begin{table}[t]
\caption[] {Equivalent width (EW), Full Width at Half Maximum (FWHM),
Full Width at Zero Intensity (FWZI) and Relative Intensity
of the major emission lines}
\begin{tabular}{lllll}
\hline
\hline\noalign{\smallskip}
Spectral line           & EW & FWHM & FWZI  & Relative \\
                & (\AA) & (km s$^{-1}$)  &  (km s$^{-1}$)  &  Intensity  \\
\noalign{\smallskip}
\hline\noalign{\smallskip}

H$_{\beta }$         & 23.2 &  950 & 3900  & 2.16 \\
H$_{\gamma }$        & 21.7 & 1000 & 3500  & 2.03 \\
H$_{\delta }$        & 14:  & 1000 & 3500: & 1.81 \\
He\,I $\lambda4471$  &  6.7 & 1100 & 2700  & 1.36 \\
He\,I $\lambda4713$  &      & 1000 &       & 1.10  \\
He\,I $\lambda4921$  &  3.4 & 1050 & 2500  & 1.18 \\
He\,II $\lambda4686$ &  6.4 & 1300 & 3500: & 1.29 \\
C\,III/N\,III $\lambda4650$ &  1.9 & 1270: &      & 1.10 \\
\noalign{\smallskip}
\hline

\end{tabular}
\label{Table1}
\end{table}

\begin{figure*}
\sidecaption
\includegraphics[width=12cm]{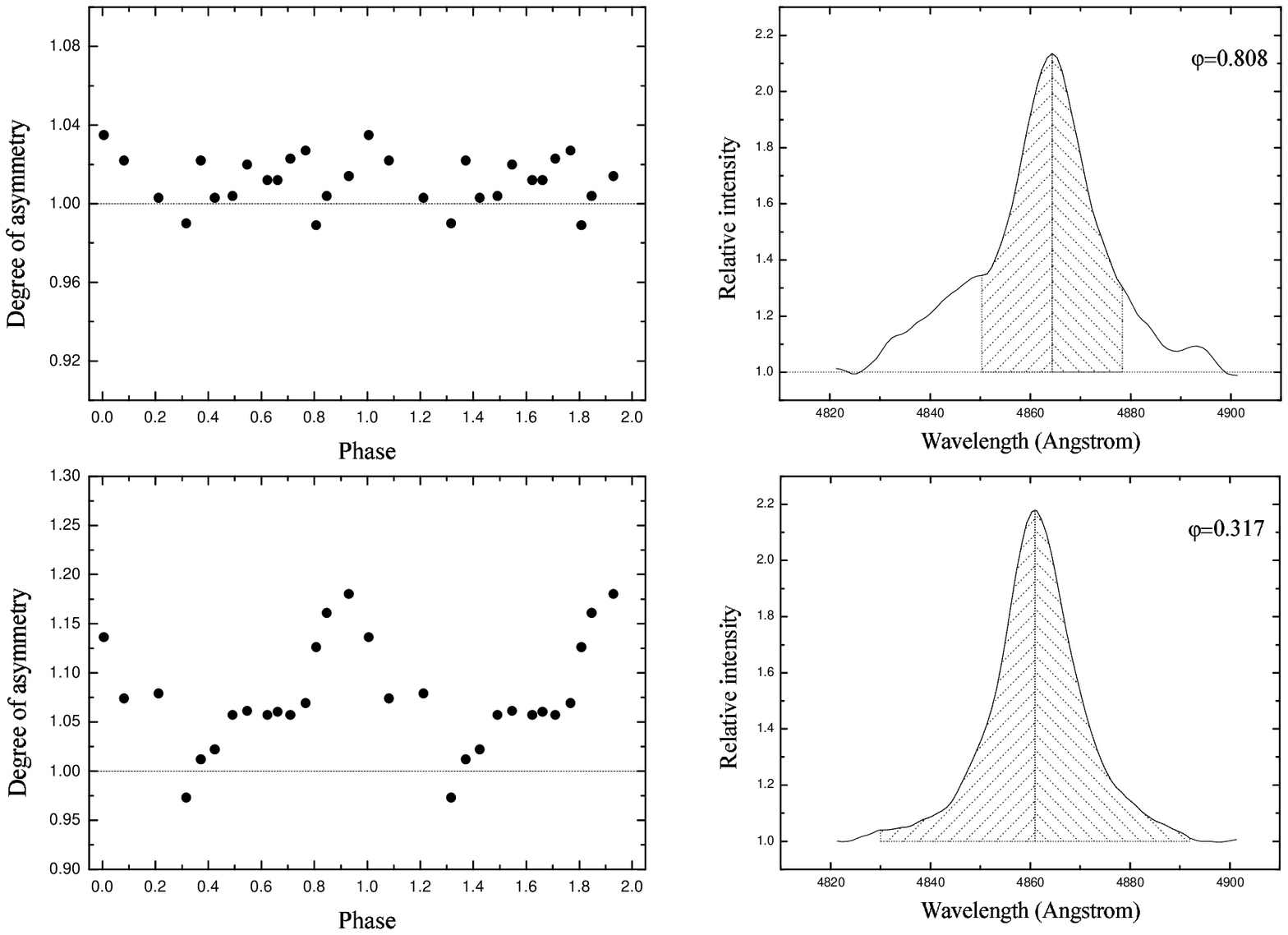}
\caption{The degree of asymmetry of the emission line H$_{\beta }$ folded on
the adopted period of 0\fd0595. The degree of asymmetry is the ratio between the 
areas of blue and red parts of the emission line. This parameter is very
similar to the V/R ratio for double-peaked emission lines.
The degree of asymmetry was calculated for two values of the wavelength range of the 
line wings (left-hand frame, upper panel for a wavelength range of 13 \AA, and lower 
panel for a wavelength range of 30 \AA). 
On the right-hand frame are shown some profiles of the H$_{\beta }$ line with the marked 
wavelength range used for determination of the degree of asymmetry.
One can see that central narrow component of line remains practically 
symmetrical throughout an orbital period, showing only slight skewness to the right from 
time to time. The broad base component of the line shows strong variability, and 
it becomes quite asymmetric at a phase of about 0.9.
}

\label{asymmetry}
\end{figure*}

\subsection{Emission lines variations}

In order to increase the signal to noise ratio of the spectra we have phased
the individual spectra with the orbital period, derived in the next section, and
then co-added the spectra into 15 separate phase bins.
Fig.~\ref{profiles} shows  the H$_{\beta }$ profiles from some of the obtained spectra.
We note a broad base component and a less narrow line component which is present 
throughout the orbital cycle. There are orbital variations in the line profiles. 
But whereas the 
narrow component remains practically symmetric throughout the orbital cycle, on the 
blue wing of the broad component at phases about 0.75 -- 0.9 a hump appears.
A similar behavior is observed for all lines, but most notably for H$_{\beta }$.

For a more accurate examination of the profiles for asymmetry we calculated the degree 
of asymmetry of all profiles as the ratio between the areas of the blue and the red parts 
of the emission line. 
This parameter is very similar to the V/R ratio for double-peaked emission lines.
It strongly depends on the wavelength range of the line wings which was selected for 
calculation of the degree of asymmetry. 
Choosing a greater wavelength range, we can analyze the farther line wings.

We calculated the degree of asymmetry for two values of the wavelength range of 
the line wings, and plotted this parameter as a function of orbital 
phase 
(left-hand frame of Fig.~\ref{asymmetry}, upper panel: for a wavelength range of 13 \AA,
and lower panel: for a wavelength range of 30 \AA). 
On the right-hand frame of Fig.~\ref{asymmetry} are shown some profiles of the H$_{\beta }$ 
line with the marked 
wavelength range used for determination of the degree of asymmetry.
One can see that the central narrow line component indeed remains practically 
symmetrical throughout the orbital period, showing only a slight skewness to the right from 
time to time (Fig.~\ref{asymmetry}, top). At the same time the broad base component 
of the line shows strong variability, and it becomes most asymmetric at a phase 
of about 0.9 (Fig.~\ref{asymmetry}, bottom).
This variability seems to contain information on the structure
of the accretion disk and give evidence for an unusually located emission
region.

\begin{figure}
\resizebox{\hsize}{!}{\includegraphics{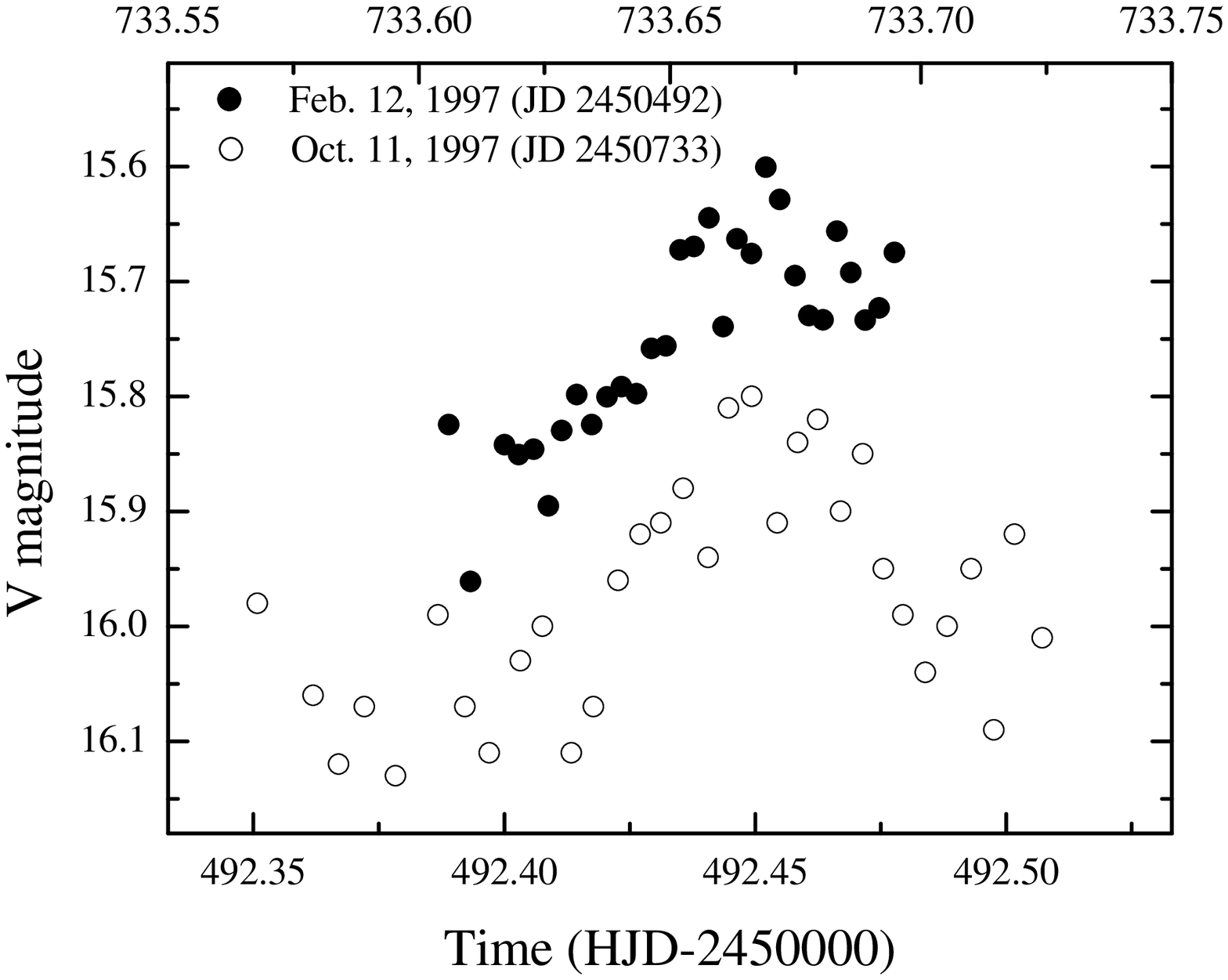}}
\caption{Light curves of FS Aur on February 12, 1997 (closed circles, bottom X axis) and 
October 11, 1997 (open circles, top X axis).}
\label{photo}
\end{figure}

\subsection{Check of the period}

Our photometric and spectral observations were carried out non-simultaneously and 
their short duration does not allow us to make any reliable period search.
However as until now the orbital period of FS Aur was based only for spectral data
(TPST), we have decided to check, if our photometric data agree with spectral 
period.
  
Both our 2.1-hour and 4-hour light curves in the V-filter (Fig.~\ref{photo}) clearly 
show medium-amplitude (0\fm3) variations, although the flickering makes
the light curves noisy. 
Although one can see a constant difference between two curves but 
their the shapes remain similar.
However no periodic modulations with spectroscopic period were detected, 
and this the period of photometric modulations should be at least 3 hours.

\begin{figure}[t]
\vspace{0.5 cm}
\resizebox{\hsize}{!}{\includegraphics{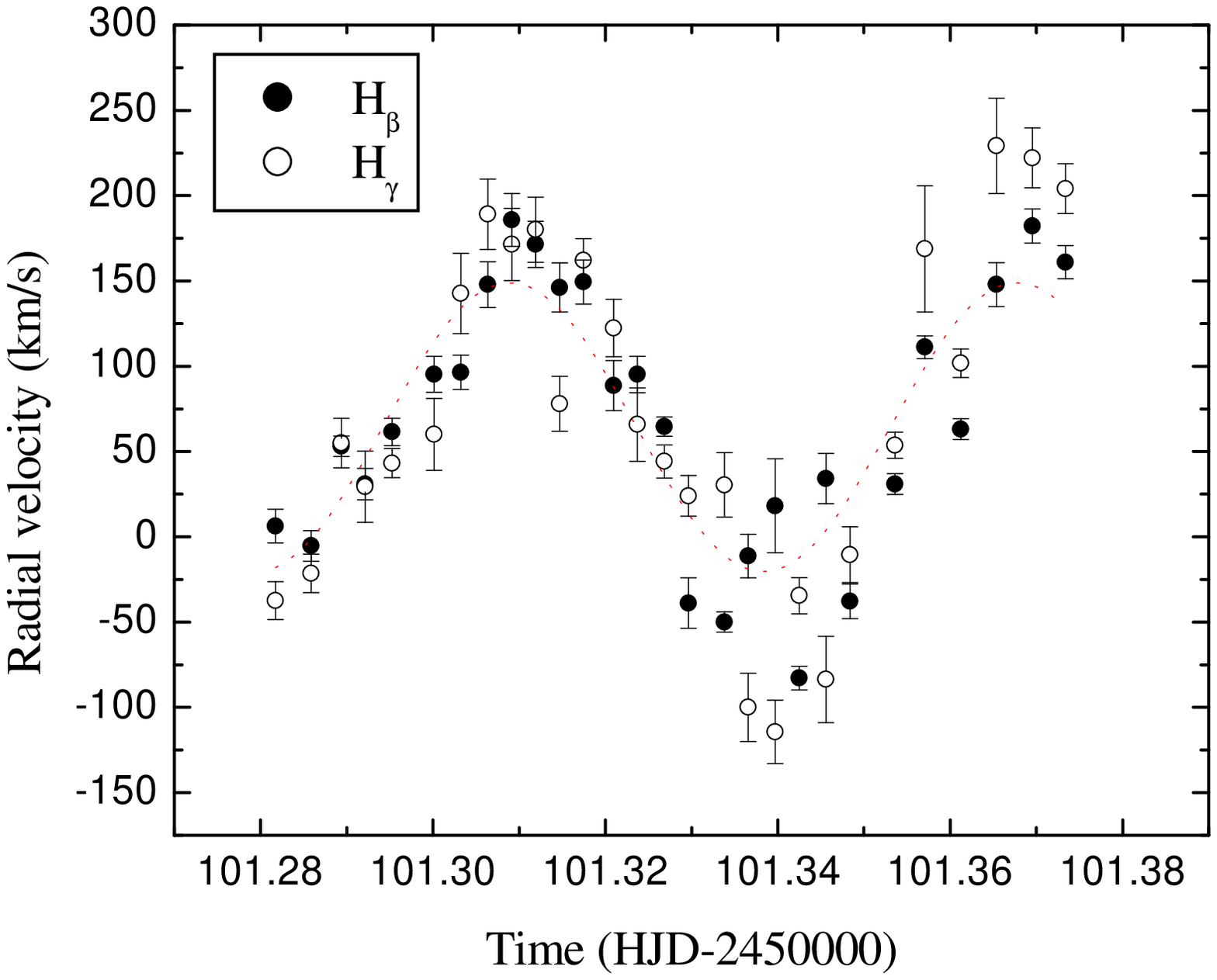}}
\caption{
The radial velocity curve of the H$_{\beta }$ and H$_{\gamma }$ emission lines,
derived from a single gaussian fitting method. To obtain an estimate of the 
period, a sine curve fit was made to the velocities, giving P=0\fd059 $\pm$ 
0\fd002 for both H$_{\beta}$ and H$_{\gamma }$
}
\label{vel_per}
\end{figure}

As we detected a discrepancy between the photometric modulation period
and the H$_{\alpha }$ velocity variations period (from TPST), we decided also to check 
the spectral period using our spectral data. To verify the orbital period we 
measured the radial velocities in the H$_{\beta }$ and H$_{\gamma }$ emission lines 
using a gaussian fit. To obtain an estimate of the period, a sine curve fit was made
to the velocities, giving P=0\fd059 $\pm$ 0\fd002 for H$_{\beta}$ and H$_{\gamma }$
(Fig.~\ref{vel_per}). 
This period agrees well with the 0\fd0595 $\pm$ 0\fd0001 estimate by TPST.
Additional evidence for this period, though less reliable, comes from the 
cyclic variations of the degree of asymmetry of the emission lines and their equivalent 
widths obtained from individual spectra, which oscillate with the same period.

Though we obtained different results based on photometric and spectral data, 
there are no reasons to doubt the reliability of the orbital (spectroscopic) period.
Previous photometry of FS Aur over a 1.8-hour time span showed a
0\fm15 modulation in the B-filter with a period between 87--105 min (Howell
\& Szkody \cite{how:szkody}) that is consistent with the spectroscopic period.
Many CVs are known, for which the photometric behaviour varies for a short time.
It is necessary to perform longer photometric observations of FS Aur during 
a single night to fully clarify this vagueness.

Note that henceforth we shall be using the period found by TPST because of 
its higher accuracy.

\begin{figure*}[t]
\resizebox{\hsize}{!}{\includegraphics{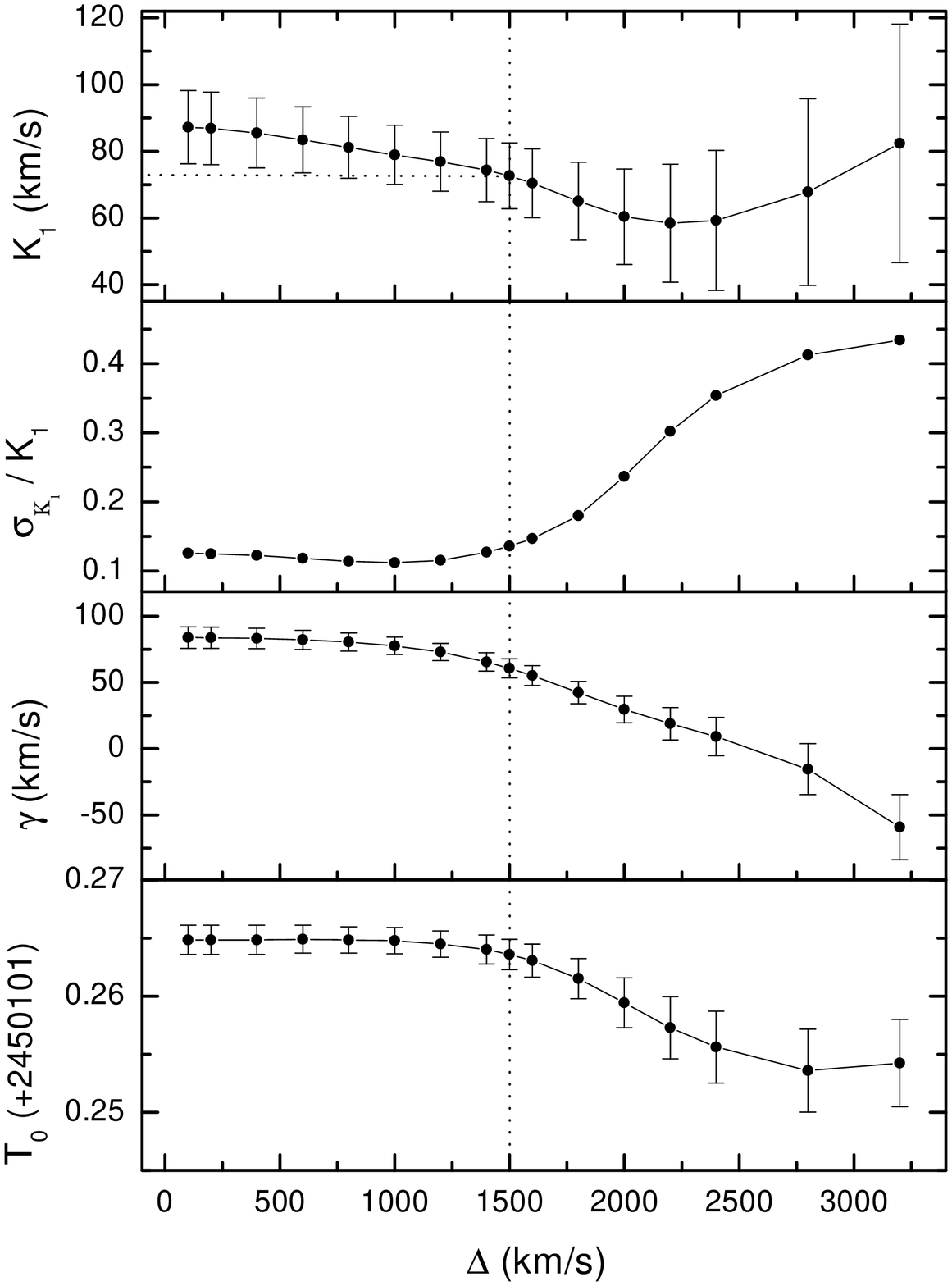}
\includegraphics{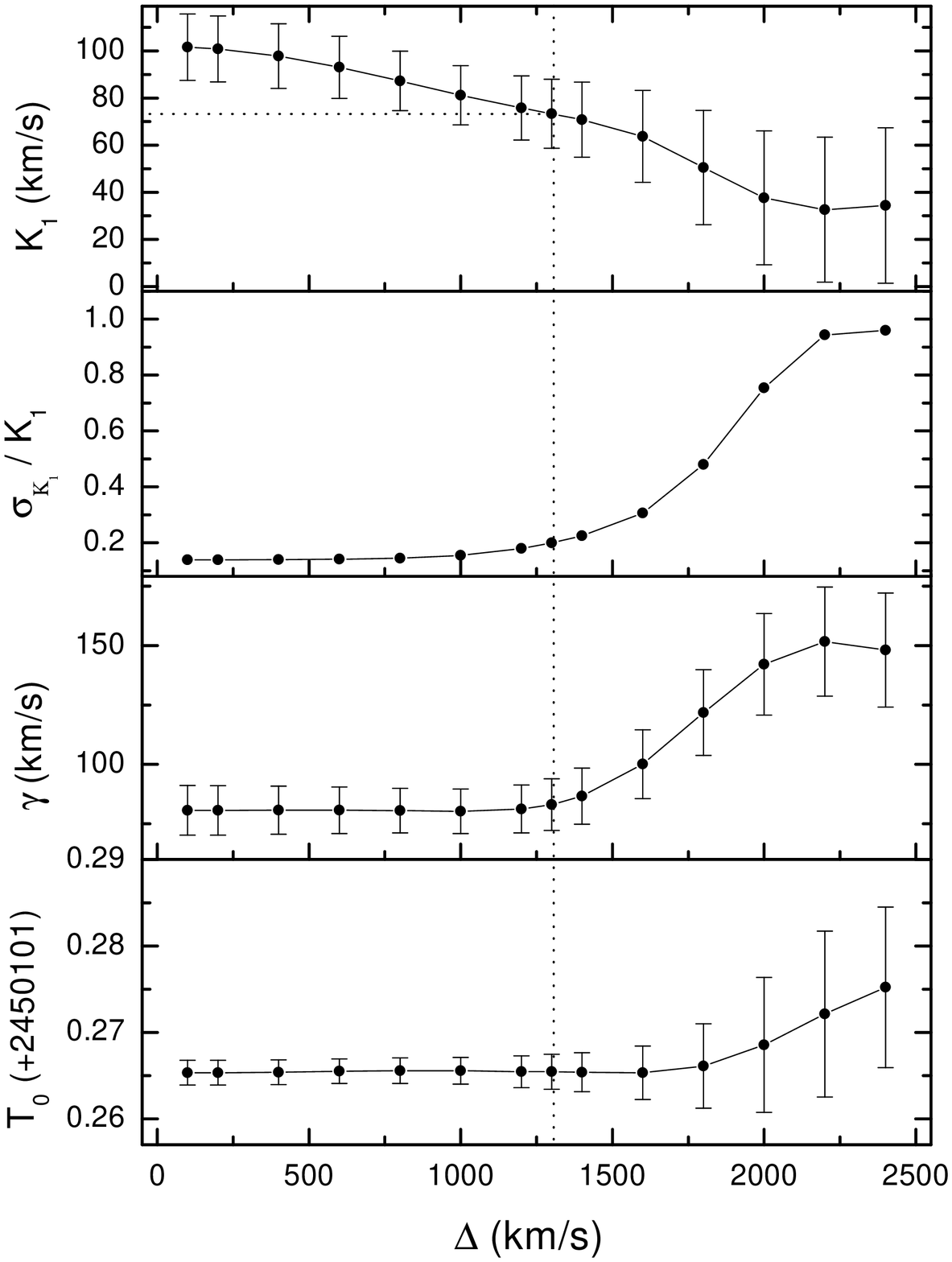}}
\caption{The diagnostic diagram for the H$_{\beta }$ (left panel) and H$_{\gamma }$ 
(right panel) data, showing the response of the fitted orbital elements to the 
choice of the double-gaussian separation. The best fit is reached with gaussian
separations of $\approx$1500 km s$^{-1}$ for H$_{\beta }$ and 
of $\approx$1300 km s$^{-1}$ for H$_{\gamma }$.}
\label{diagram}
\end{figure*}

\subsection{The radial velocity curve}
\label{radial}

In cataclysmic variables the most reliable parts of the emission line profile 
for deriving the radial velocity curve are the extreme wings. They
are presumably formed in the inner parts of the accretion disk and therefore
should represent the motion of the white dwarf with the highest reliability.

The velocities of the emission lines were measured using the double-gaussian method
described by Schneider \& Young (\cite{sch:young}) and later refined by
Shafter (\cite{shafter1}).
This method consists of convolving each spectrum with a pair of gaussians of
width $\sigma $ whose centers have a separation of $\Delta $. 
The position at which the
intensities through the two gaussians become equal is a measure of the
wavelength of the emission line. The measured velocities will depend on the
choice of $\sigma $ and $\Delta $, and by varying $\Delta $ different
parts of the lines can be sampled. The width of the gaussians $\sigma $ is
typically set by the resolution of the data.

We have measured the velocities in our binned spectra for the four emission 
lines (H$_\beta ,$ H$%
_\gamma $, He\,I $\lambda $4471 and He\,II $\lambda $4686) separately in order
to test for consistency in the derived velocities. We measured the radial
velocity using gaussian separations ranging from 50 km s$^{-1}$ to 3200 km s$%
^{-1}$. All measurements were made using $\sigma =200$ km s$^{-1}$ and
$\sigma =300$ km s$^{-1}$. For each value of $\Delta $ we made a
non-linear least-squares fit of the derived velocities to sinusoids of the
form


\begin{figure}[t]
\resizebox{\hsize}{!}{\includegraphics{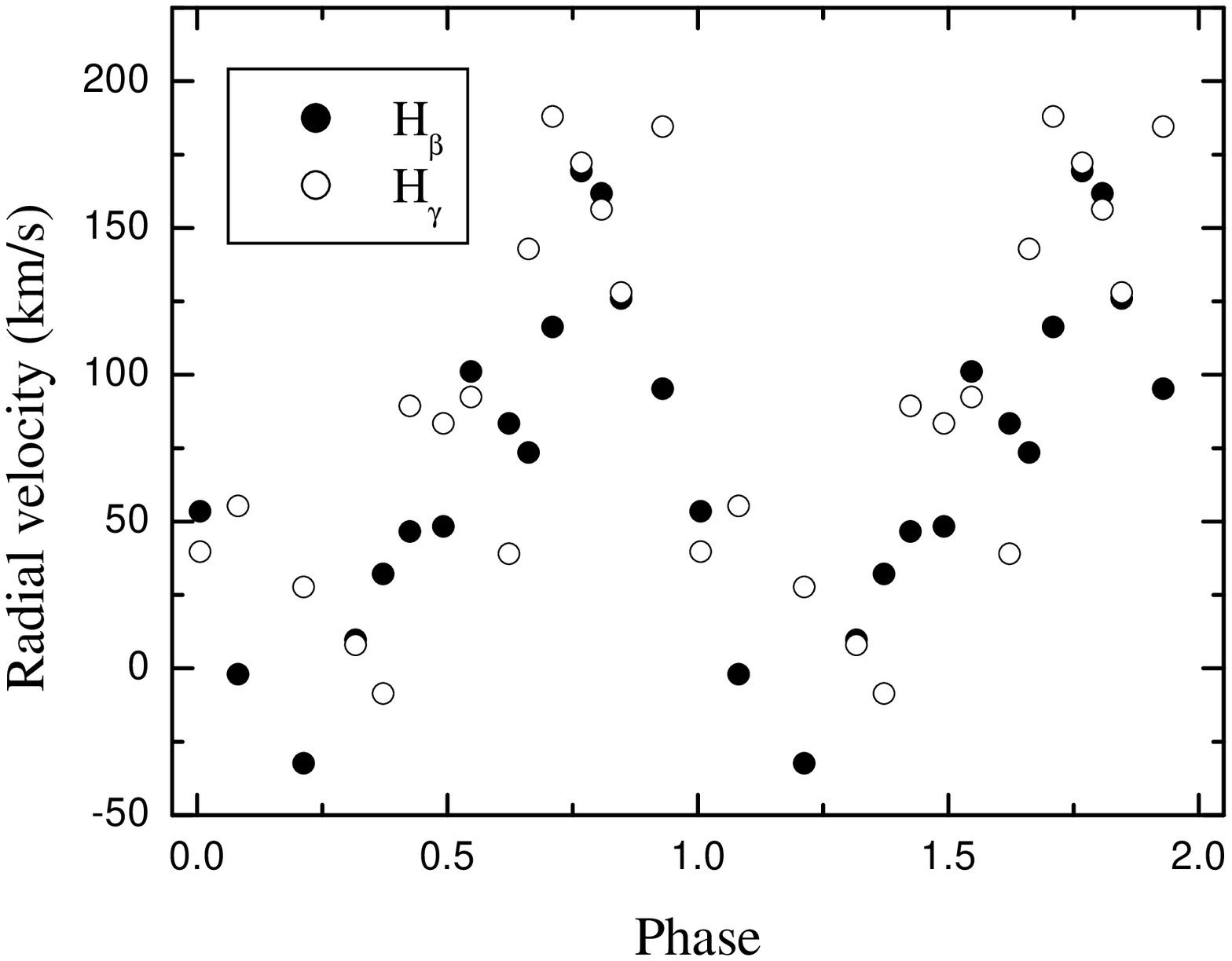}}
\caption{The H$_{\beta }$ (closed circles) and H$_{\gamma }$ (open circles)
radial velocities measured using the double-gaussian method folded on the 
orbital period. All data are plotted twice for continuity. The K-velocity is
73$\pm$10 km s$^{-1}$ and 73$\pm$15 km s$^{-1}$ respectively.}
\label{velocity}
\end{figure}


\begin{equation}  \label{squaresfit}
V(t,\Delta )=\gamma (\Delta )+K(\Delta )\sin \left[ 2\pi \left( T-T_0\left(
\Delta \right) \right) /P+\pi\right]
\end{equation}
where $\gamma $ is the systemic velocity, $K$ is the semi-amplitude, $T_0$ is
the time of inferior conjunction of the secondary star and $P$ is the orbital period. The
resulting ``diagnostic diagram'' for H$_\beta $ and H$_\gamma $ with
$\sigma=200$ km s$^{-1}$ is shown in Fig.~\ref{diagram}. The diagram shows the
variations of $K$, $\sigma(K)/K$ (the fractional error in $K$), $\gamma $ and
$T_0$ with $\Delta $ (Shafter et al. \cite{shafter3}).
The diagram for $\sigma =300$ km s$^{-1}$ looks the same.

To derive the orbital elements of the line wings we took the values that
correspond to the largest separation just before $\sigma(K)/K$ shows a
sharp increase (Shafter \& Szkody \cite{shafter2}).
Note that the dependence of parameter $\sigma(K)/K$ on Gaussian separation 
for both hydrogen lines is very similar (Fig.~\ref{diagram}). For H$_\beta $ it
appears that $\Delta $ can be increased to $\sim $1500 km s$^{-1}$ before
$\sigma(K)/K$ begins to increase. Similarly, the optimum value of $\Delta $ for H$_\gamma $ 
probably lies near 1300 km s$^{-1}$.
Referring to the diagnostic diagram, the
$K$ values for H$_\beta $, H$_\gamma $, He I $\lambda $4471 and He II
$\lambda$4686 are 73, 73, 81 and 69 km s$^{-1}$, respectively.
The measured parameters of the best fitting radial velocity curves 
are summarized in Table~\ref{Table3}. In Fig. \ref{velocity} 
we show the radial velocity curves of H$_\beta $ and H$_\gamma $ emission lines.

\begin{table}
\caption[]{Elements of the radial velocity curves of FS Aur}
\begin{flushleft}
\begin{tabular}{cccc}
\hline\hline
\noalign{\smallskip}
Emission & $\gamma$-velocity & K$_{1}$ & T$_{0}$ \\
line & (km s$^{-1}$) & (km s$^{-1}$) & (HJD) \\

\noalign{\smallskip}
\hline
\noalign{\smallskip}
H$_{\beta }$ & 46$\pm$7 & 73$\pm$10 & 2450101.264$\pm$0.001 \\
H$_{\gamma}$& 68$\pm$11 & 73$\pm$15 & 2450101.265$\pm$0.001 \\
He\,I $\lambda$4471& 113$\pm$10 & 81$\pm$15 & 2450101.263$\pm$0.002 \\
He\,II $\lambda$4686 & 50$\pm$21 & 69$\pm$29 & 2450101.265$\pm$0.004 \\
\noalign{\smallskip}
\hline
\noalign{\smallskip}
\textbf{Mean} & \textbf{52$\pm$7} & \textbf{73$\pm$8} & \textbf{2450101.264$\pm$0.001} \\

\noalign{\smallskip}
\hline

\end{tabular}
\label{Table3}
\end{flushleft}
\end{table}

Basically, the radial velocity semi-amplitudes of the Balmer and Helium
lines are consistent, while the $\gamma $-velocities are
not. The reason for this is unknown. In the discussion to follow we will
adopt a mean value of $K$ and $T_{0}$ for these lines
(using the $\sigma(K)$ and $\sigma(T_{0})$ as a weight factor): 
$K$=73$\pm$8 km s$^{-1}$ and $T_{0}(HJD)$=2450101.264$\pm$0.001.

\begin{figure}
\resizebox{\hsize}{!}{\includegraphics{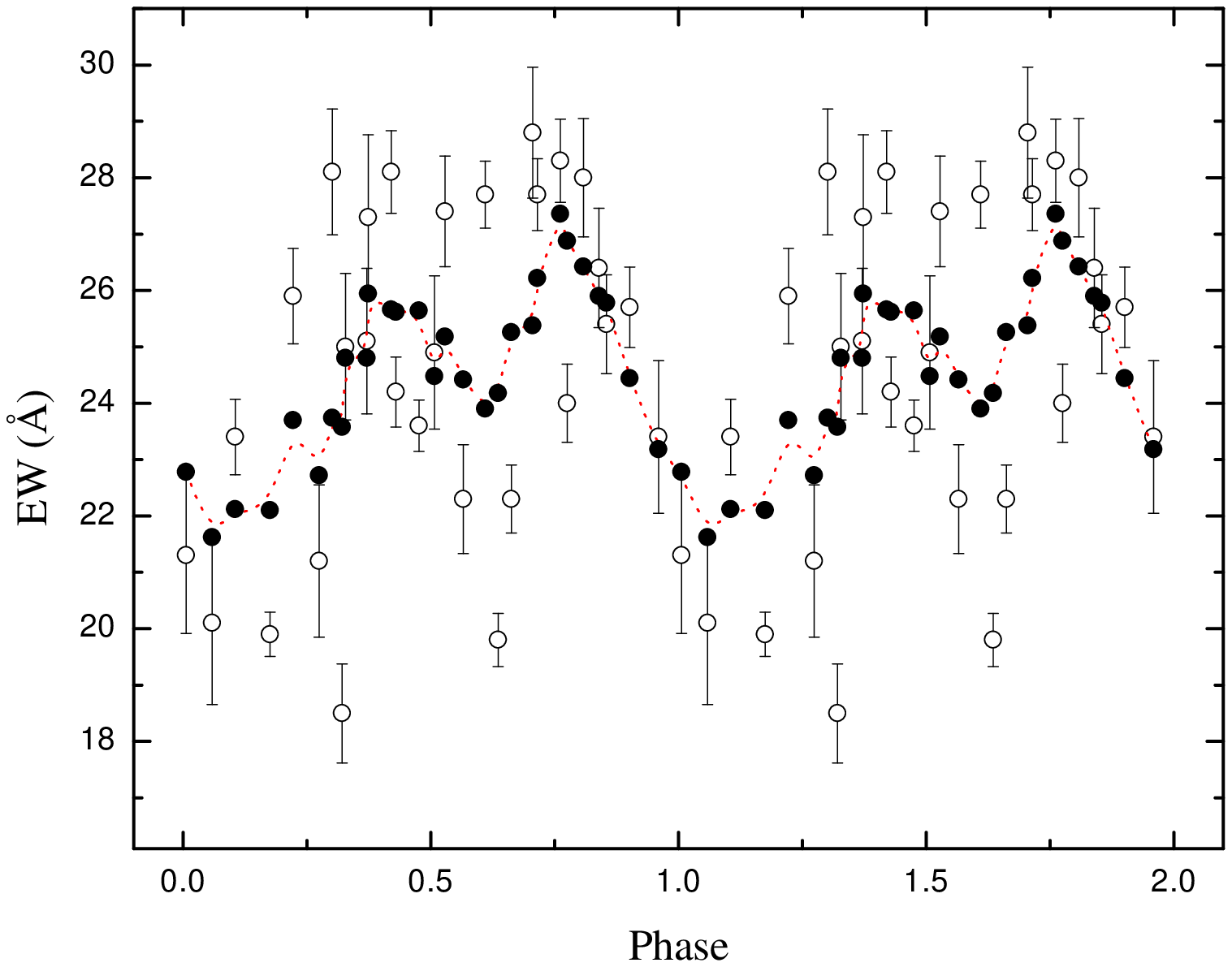}}
\caption{The variation of the H$_{\beta }$ equivalent widths with orbital period.
Open circles shows individual values, filled circles with dotted line show data 
which were obtained by averaging adjacent data points.
The data are plotted twice for continuity.}
\label{eqw}
\end{figure}

\subsection{Equivalent width}

We measured the equivalent widths of the H$_{\beta }$ emission line in all 
individual spectra. They were investigated for a modulation with orbital 
period (Fig.~\ref{eqw}). The errors have been obtained by calculating 
standard deviations from several independent measurements of the same lines. 

Though the obtained equivalent widths exhibit rather significant 
dispersion, none the less one can see their modulation with 
orbital period. It is especially a noticeable on the smoothed graph 
which was obtained by averaging adjacent data points (filled circles 
with dotted line on 
Fig.~\ref{eqw}). We found that the EW is modulated with amplitudes not 
less than 25\% of the mean value. We can confidently assert that there 
is a broad minimum in the EW around phase 0.1, and probably there is a 
secondary minimum near phase 0.6.

The observed minima could be due to an increase in the continuum luminosity
when the enhanced emission region crosses the line-of-sight. If this is so 
then Fig.~\ref{eqw} testifies about complex and unusual accretion structure 
in FS Aur.


\subsection{Possible system's parameters}
\label{syspar}

It is impossible to measure the components masses and the orbital inclination
in a non-eclipsing, single-lined spectroscopic binary like FS Aur. However,
we can determine preliminary values for the basic system's parameters, using
the assumption that the secondary is a zero-age main-sequence (ZAMS) star
(Patterson \cite{patterson}) and that the secondary fills its Roche lobe. First of all,
from our spectroscopic and photometric data we can limit the range of
possible solutions. 

Our observations reveal no evidence for eclipses, so we
expected the inclination to be less than 65$^{\circ }$. Now, from the mass
function for the system:

\begin{equation}
\frac{M_{2}\sin ^{3}i}{(1+q^{-1})^{2}}=\frac{PK^{3}}{2\pi G}
\label{MassFunction}
\end{equation}
\vspace*{0.3cm}

\noindent and any mass-period relations for the secondary stars we can obtain 
a lower limit to the mass ratio.
At using of any of the recently obtained empirical relations 
(see, for example,  Caillault \& Patterson \cite{cail:patt}, Warner \cite{warner95}, 
Smith \& Dhillon \cite{smith:dhillon}) we obtain a mass for the secondary less 
than $0.1 M_{\sun}$. Now, using Eq.(\ref{MassFunction}) we find $q>0.22$ and,
consequently, $M_{1}<0.46 M_{\sun}$. 

Also we can place a stringent lower limit on $M_{1}$, assuming that the
largest velocity in the emission line profile ($FWZI/2$) does not exceed the
Keplerian velocity at the surface of the white dwarf:

\vspace*{0.5cm}
\begin{equation}
\left( \frac{FWZI}{2\sin i}\right) ^{2}\frac{R_{1}}{G}<M_{1}  \label{FWZI}
\end{equation}
\vspace*{0.3cm}

\noindent $FWZI$ is not a well defined quantity because of the difficulty
in establishing where the high velocity line wings end and the continuum begins.
Besides, extensive wings due to Stark broadening can produce a spurious
enhancement of emissivity at small radii. However, for the time being we
assume that Stark broadening is not significant. The Balmer data 
show that the line wings extend to at least 1750 km s$^{-1}$
from the line center. 
Using $FWZI/2=1750$ km s$^{-1}$, inclination angle $i<65^{\circ }$ and
the mass-radius relation for white dwarfs (Hamada \& Salpeter \cite{hamada:salpeter}),

\begin{equation}
\frac{{R_1 }}{{R_{\sun}}} = 0.0072\left( {\frac{{M_1 }}{{M_{\sun}}}} \right)^{ - 0.8}, 
\label{MassWD}
\end{equation}
\vspace*{0.3cm}
\noindent we obtained $M_{1}>0.34 M_{\sun}.$
And finally, using the last constraint on $M_{1}$ and Eq.(\ref{MassFunction}), 
we can obtain a lower limit to the inclination: $i>51^{\circ }$.
Table \ref{Syspar} summarizes all our calculated parameters.

\begin{table}
\caption[] {Adopted system's parameters for FS Aurigae}
\begin{tabular}{ll}
\hline
\hline\noalign{\smallskip}
Parameter & Value \\
\noalign{\smallskip}
\hline\noalign{\smallskip}

T$_{0}$ (Spectroscopic Phase 0.0)   & 2450101.264$\pm$0.001 \\
Primary mass $M_{1}$ (M$_{\sun}$)                   &  0.34 -- 0.46 \\
Secondary mass $M_{2}$ (M$_{\sun}$)                 &  $\leq$0.1 \\
Mass ratio $q=M_{2}/M_{1}$                          &  $\geq$0.22 \\
Inclination $i$                                     &  $51^{\circ}$-- $65^{\circ}$ \\
\noalign{\smallskip}
\hline

\end{tabular}
\label{Syspar}
\end{table}

\begin{figure*}
\centering
\resizebox{17cm}{!}{
\includegraphics{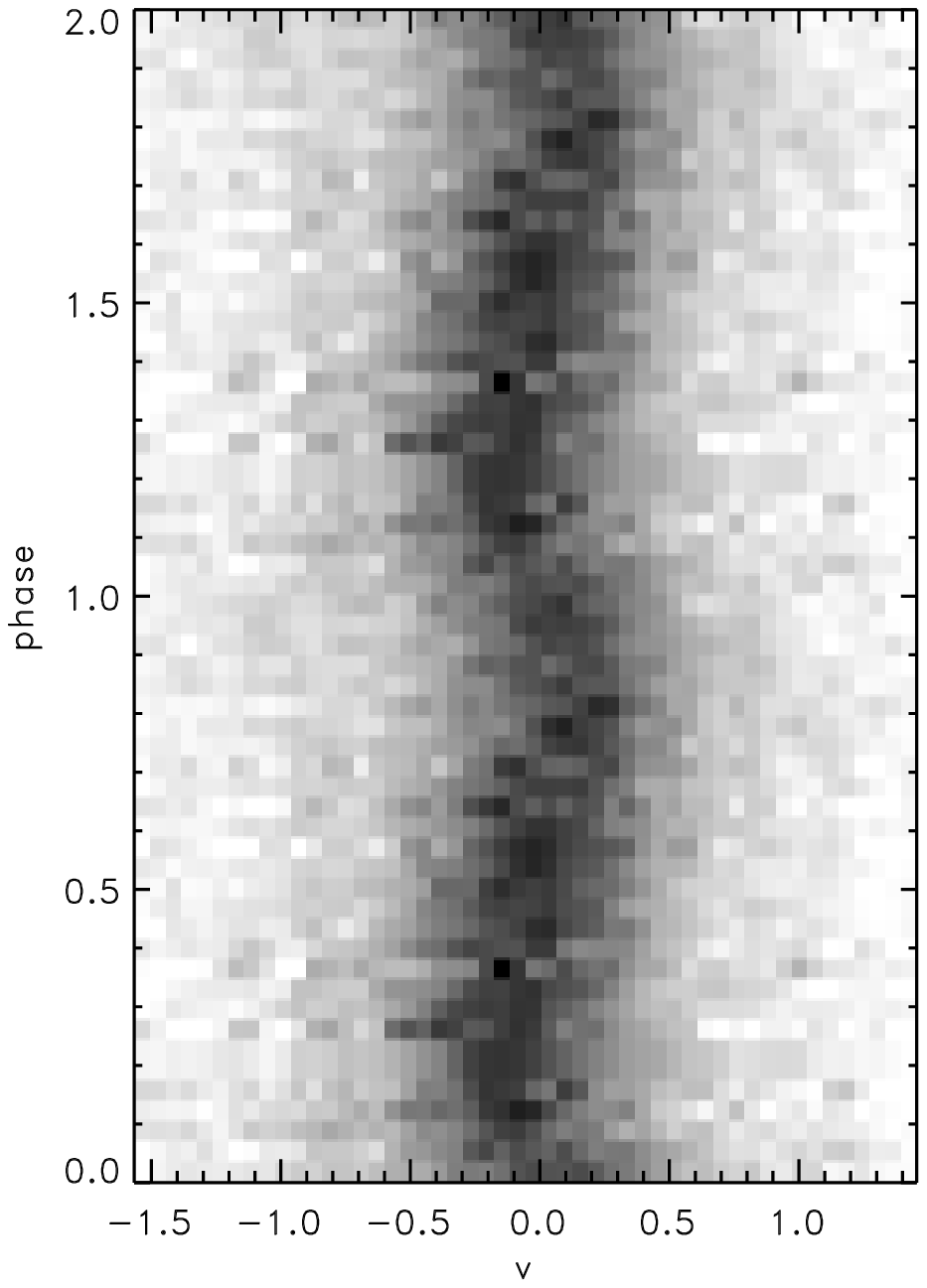}
\includegraphics{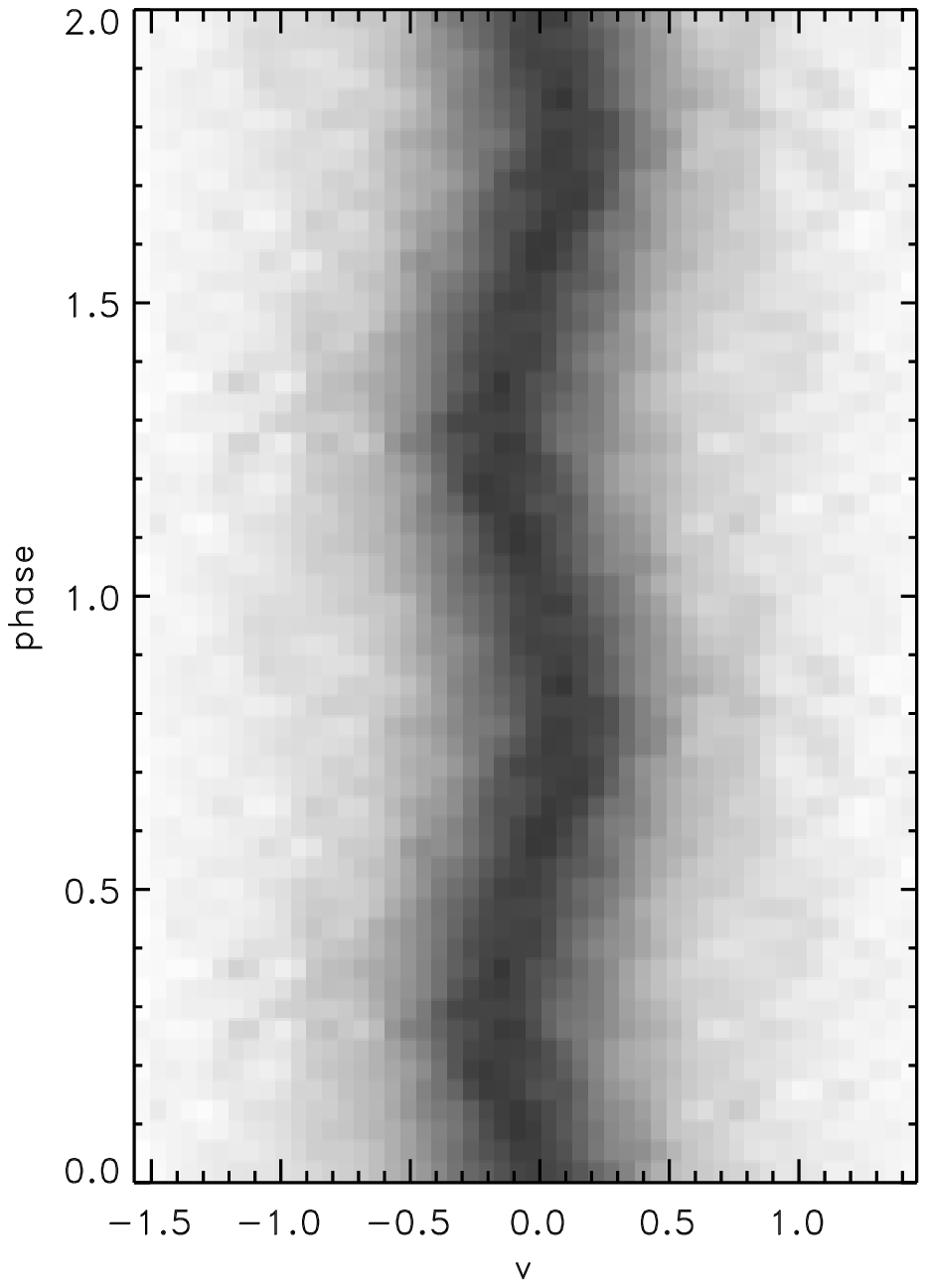}
\includegraphics{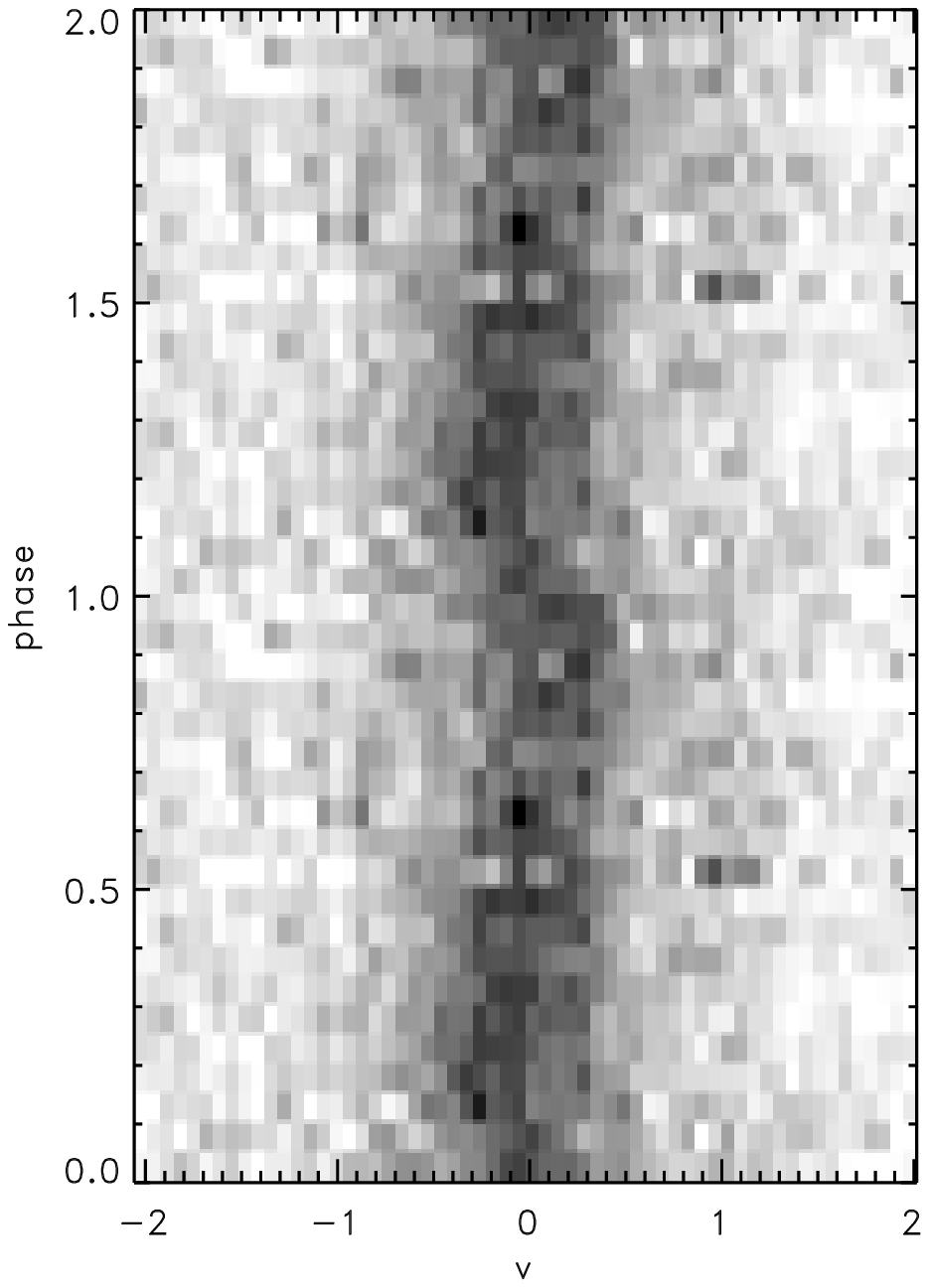}
\includegraphics{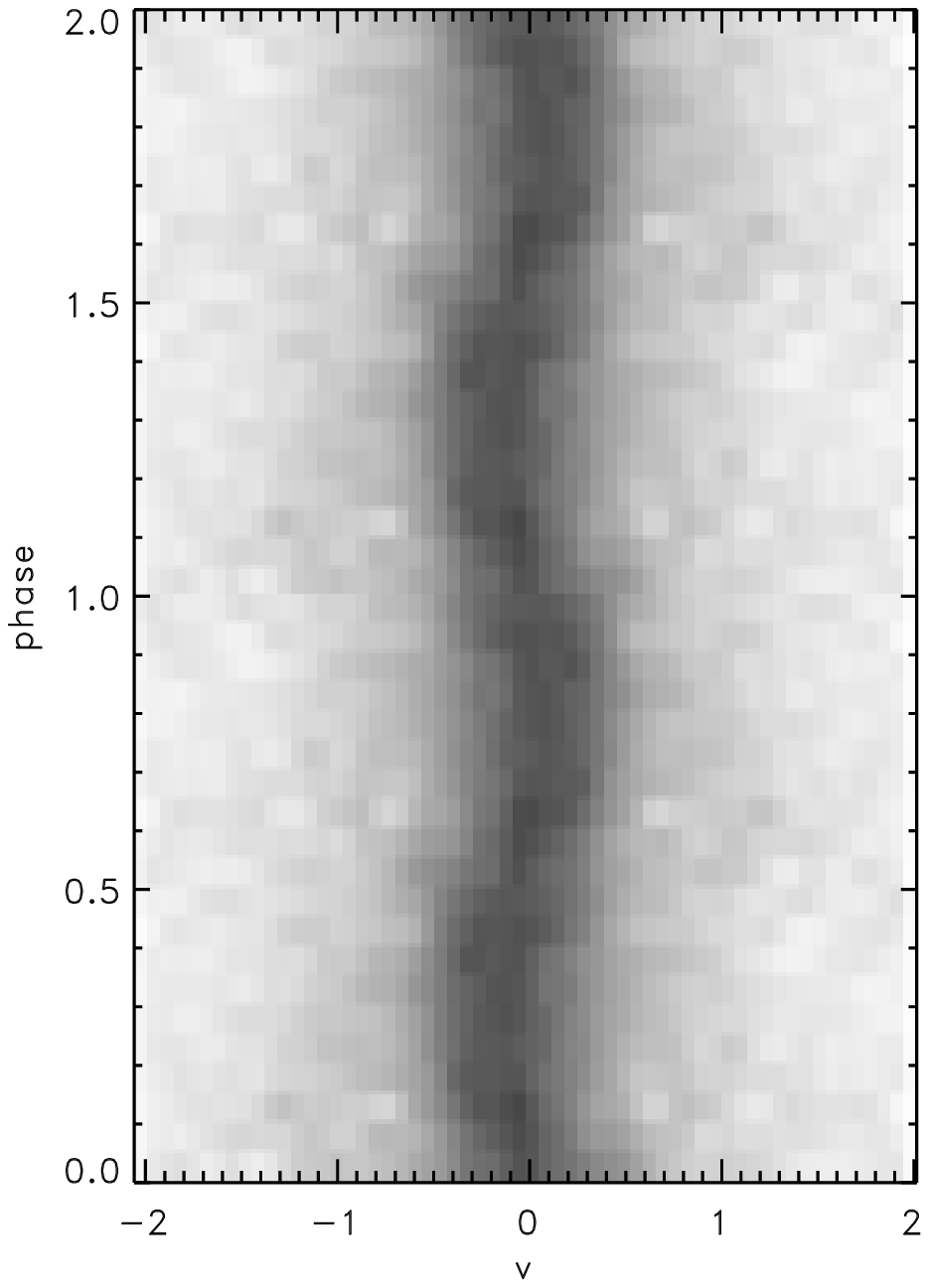}
}
\hfill
\\
\resizebox{17cm}{!}{
\includegraphics{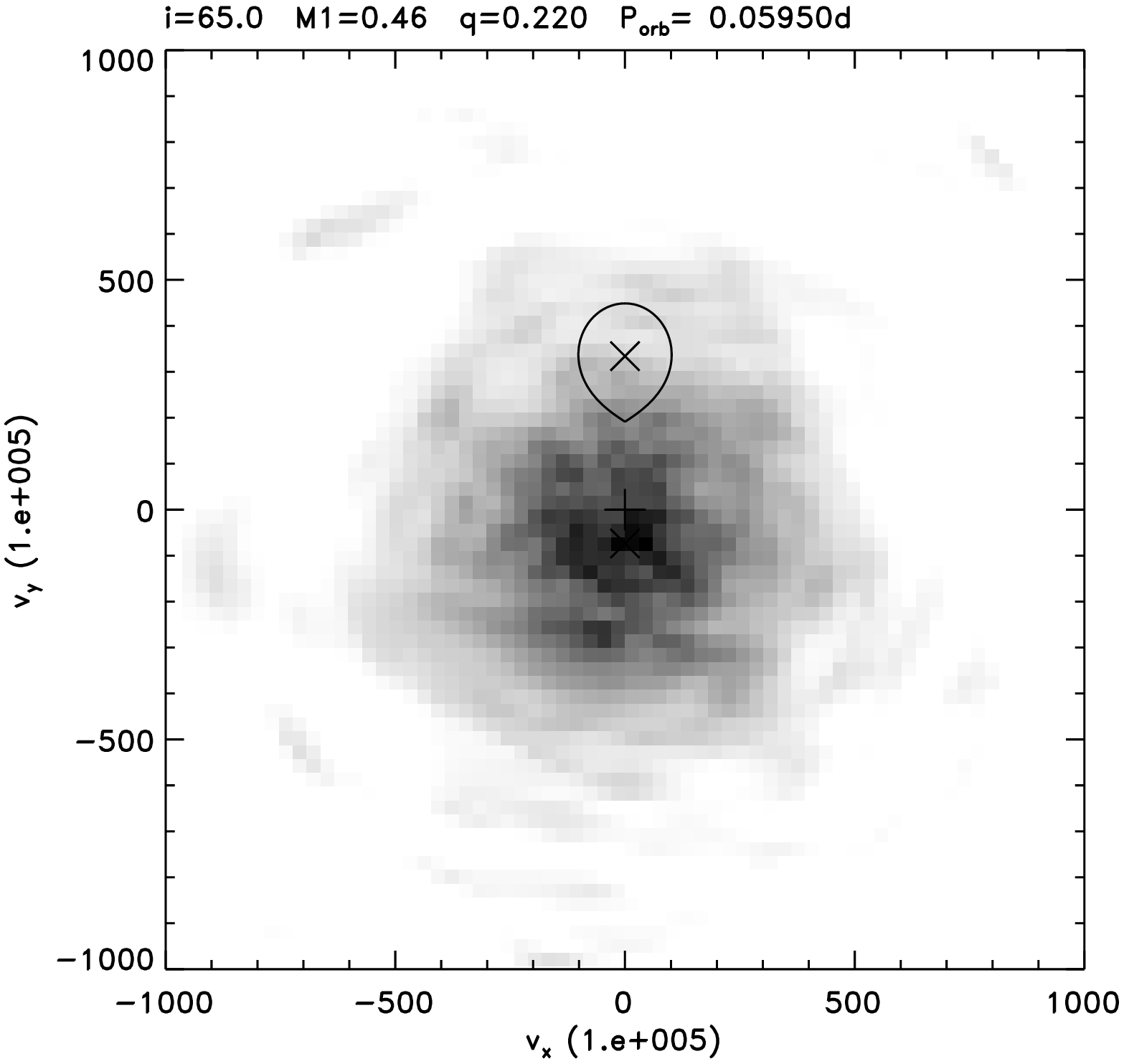}
\includegraphics{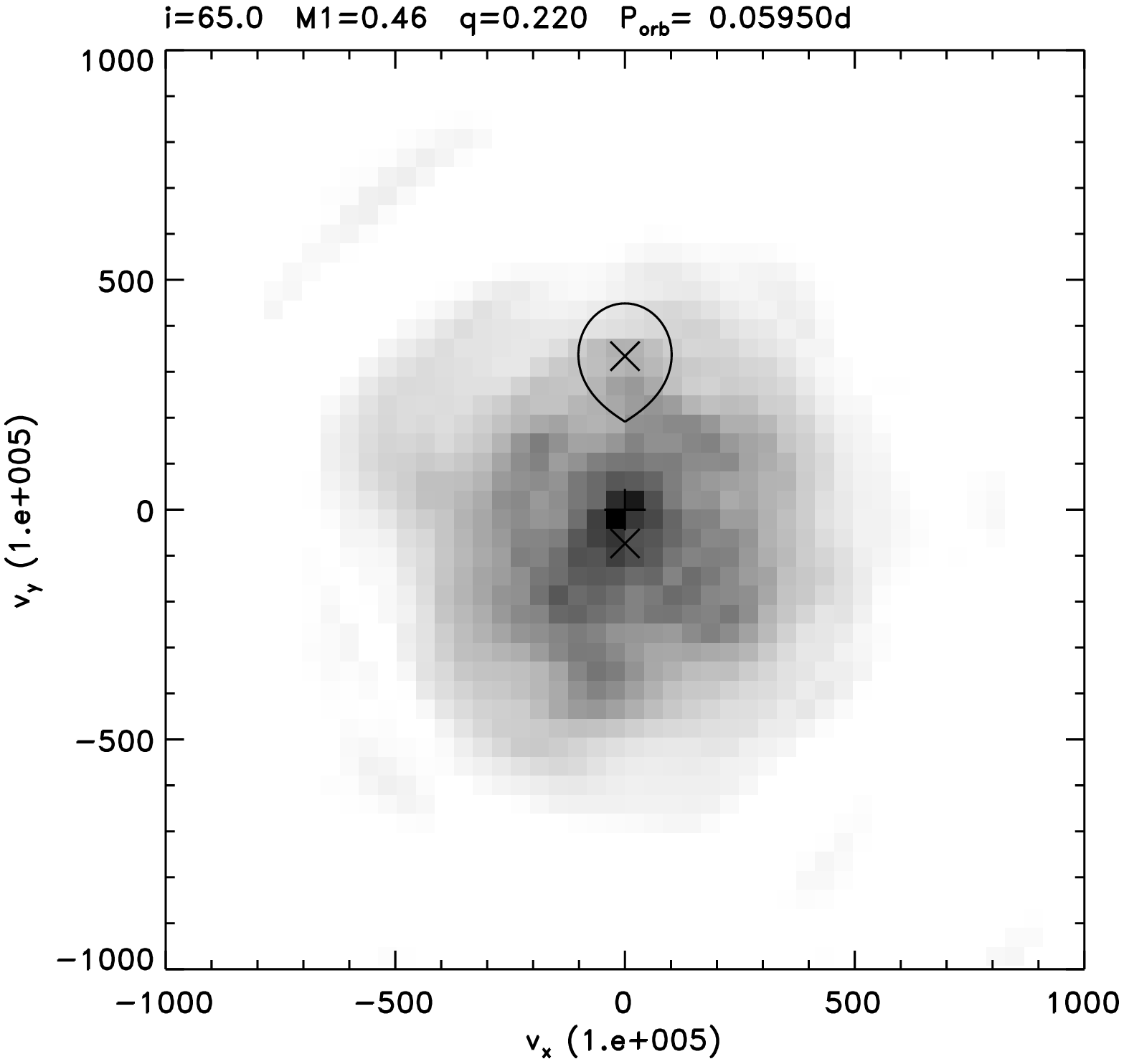}
}
\hfill
\vspace*{0.5cm}
\caption{Doppler tomograms for the H$_{\beta }$ (left) and H$_{\gamma }$ 
(right) emission lines of FS Aur. These maps display a roughly symmetric and 
very nonuniform distribution of the emission centered near to the white dwarf.
In addition to the symmetric distributed emission, at least two additional 
emission sources can be seen. The first, brighter enhanced emission component 
is roughly centered at (V$_x \approx -100$ km s$^{-1}$, V$_y \approx -260$ km s$^{-1}$)
(the appropriate phase of the cross of the line-of-sight to this bright 
region is about 0.6). The second source is located opposite the first one and 
occupies an extensive area at phases about $0.85 - 1.15$. In addition, the tomogram 
of H$_\gamma $ shows an emission ring with a radius of about 
225 km s$^{-1}$, which is centered at (V$_x \approx 0$ km s$^{-1}$, 
V$_y \approx -20$ km s$^{-1}$). In the center of this ring there is a compact bright 
spot.}
\label{top_bg}
\end{figure*}


\section{Doppler tomography}

The orbital variation of the emission lines profiles detected by us indicates a 
non-uniform structure of the accretion disk. The distribution of the disk's emission
can be explored by computing a Doppler map, using the method of Doppler tomography. 
Doppler tomography is an indirect imaging technique which can be used to determine 
the velocity-space distribution of the emission in close binary systems.
A tomogram is constructed from the line profiles obtained at a variety of orbital
phases. In other words, the Doppler map accumulates information about all profiles 
of the emission line in different phases of an orbital period.
An accretion disk producing usually double-peaked emission lines should 
appear on the tomogram as a ring with an inner radius of $V_{in} \sim 600 - 800$ 
km s$^{-1}$, plus additional emission that can be seen extending outward from 
the ring to a velocity of over $V_{out} \sim 1200$ km s$^{-1}$ or more 
corresponding to the rest of the disk. This is because the outer edge of 
the disk becomes the inner edge 
in velocity coordinates, while the inner disk is represented by the outermost 
parts of the image. But we can also obtain single-peaked lines from the accretion disk.
This can happen for many reasons, for example insufficient spectral resolution, 
a small inclination angle of the binary system, some line-broadening mechanisms. 
In this case the distribution of the emission on tomograms will not be ring-shaped 
but circular.
Full technical details of the method are given by Marsh \& Horne (\cite{marsh:horne})
and Marsh (\cite{marsh2001}). Examples of the application of Doppler tomography
to real data are given by Marsh (\cite{marsh2001}). 

\begin{figure*}
\centering
\resizebox{17cm}{!}{
\includegraphics{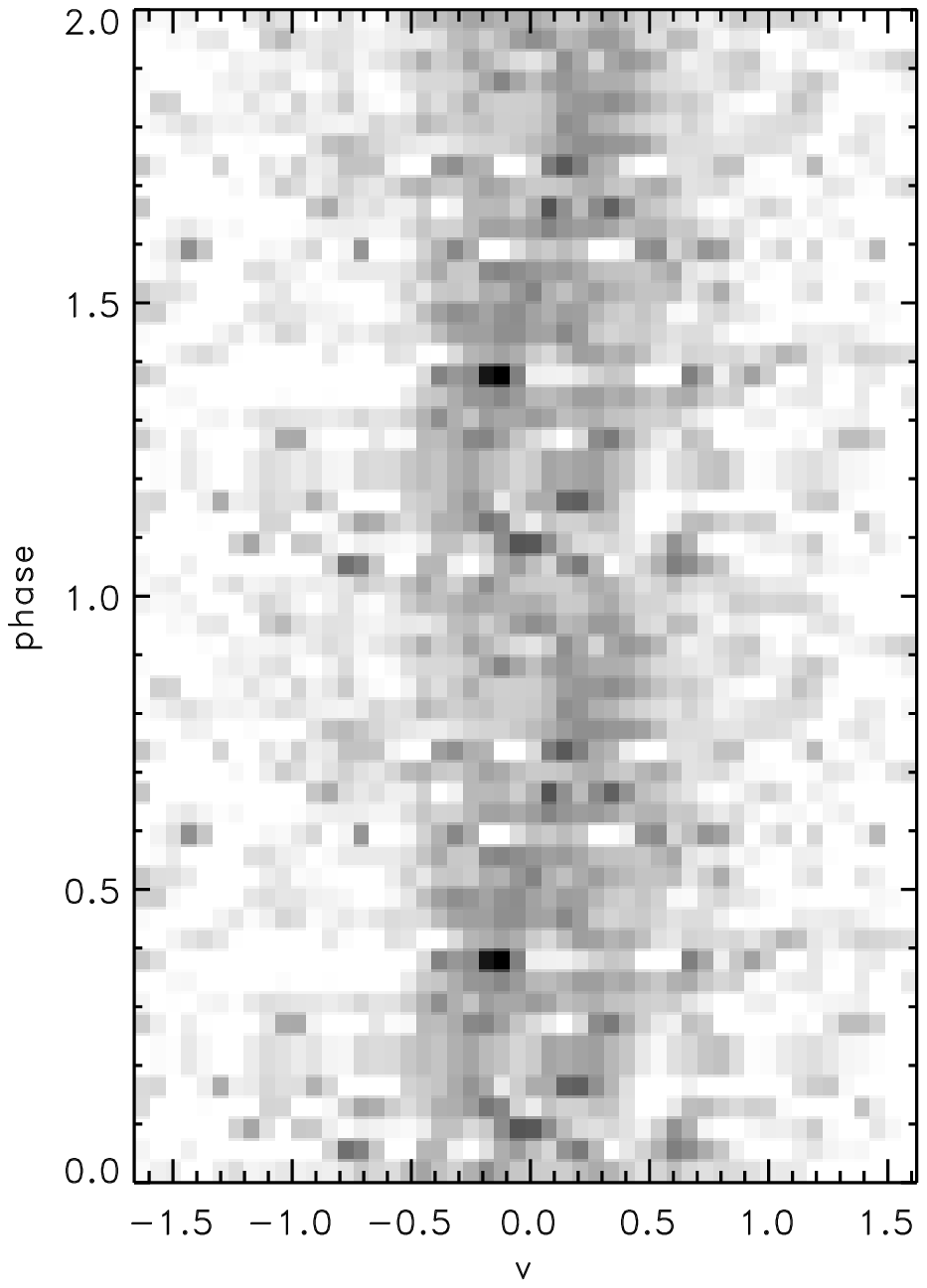}
\includegraphics{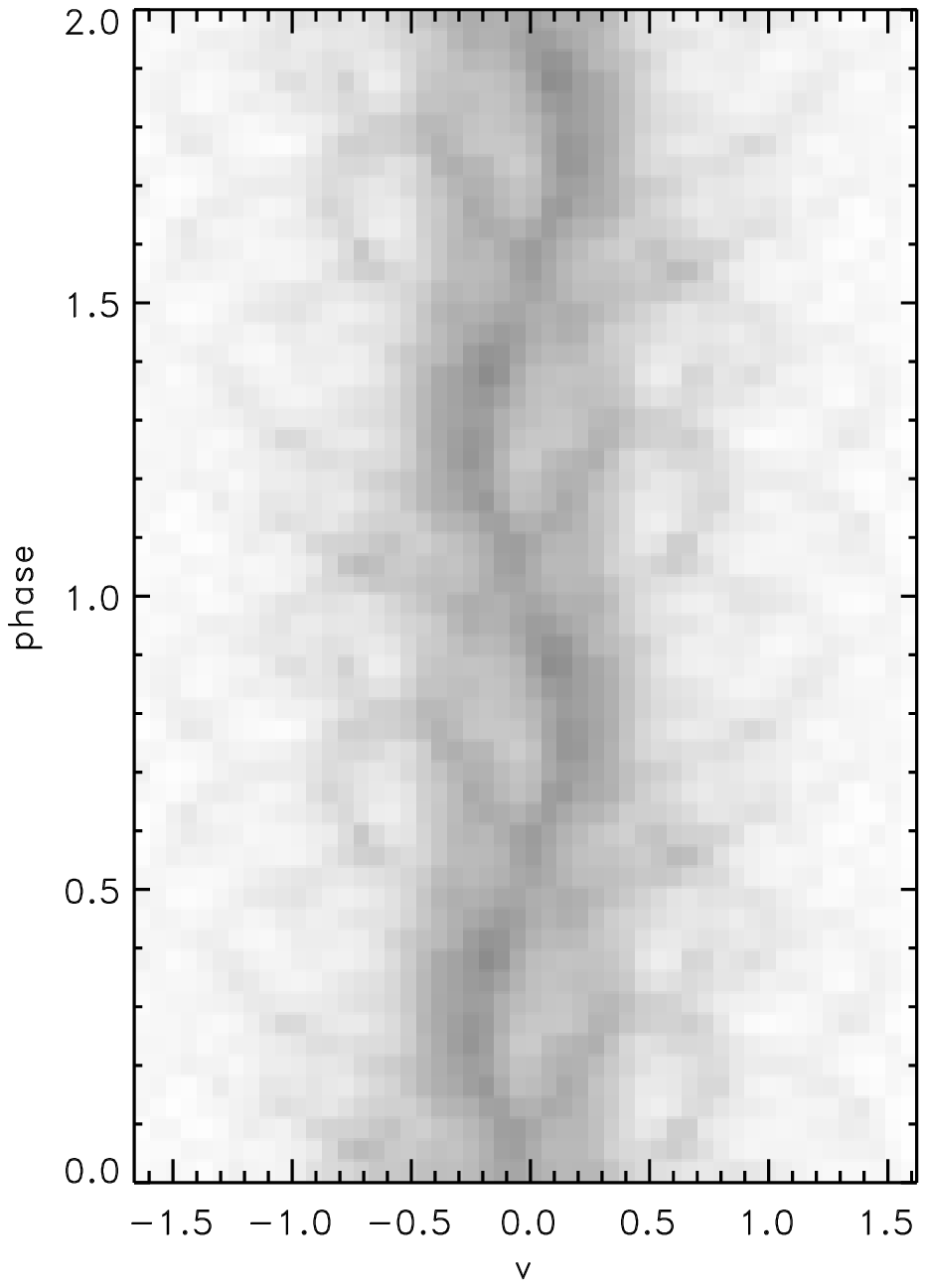}
\includegraphics{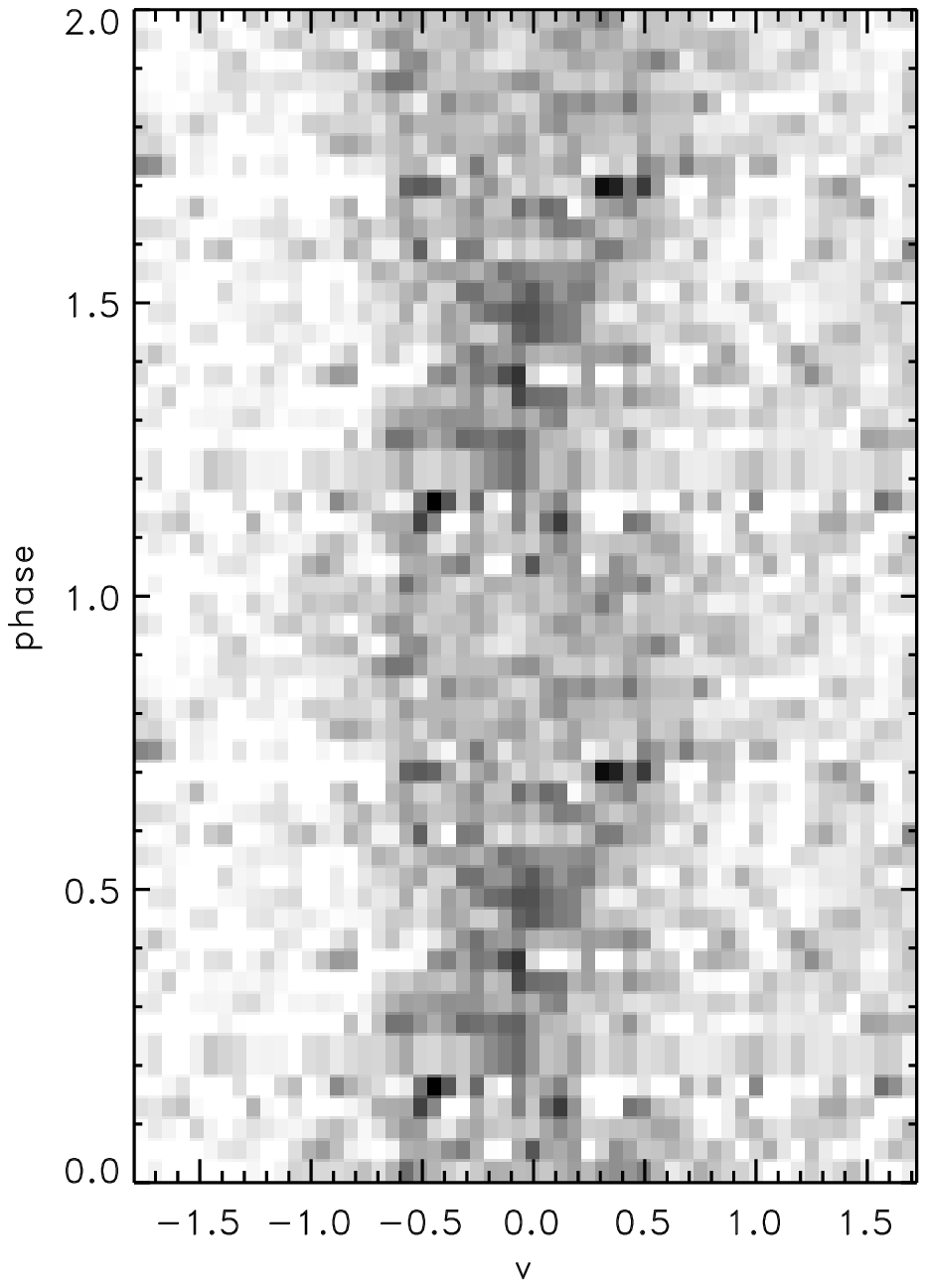}
\includegraphics{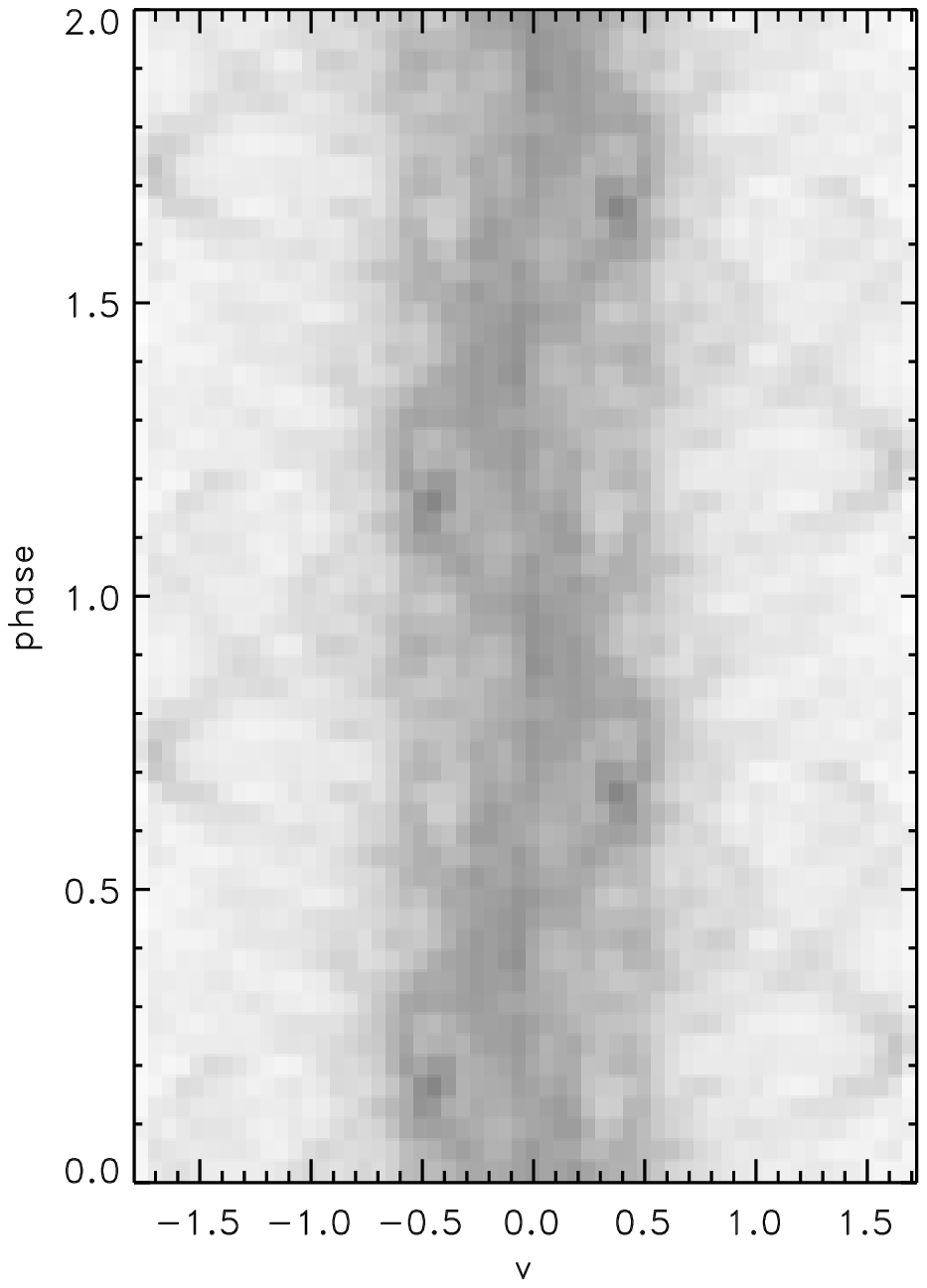}
}
\hfill
\\
\resizebox{17cm}{!}{
\includegraphics{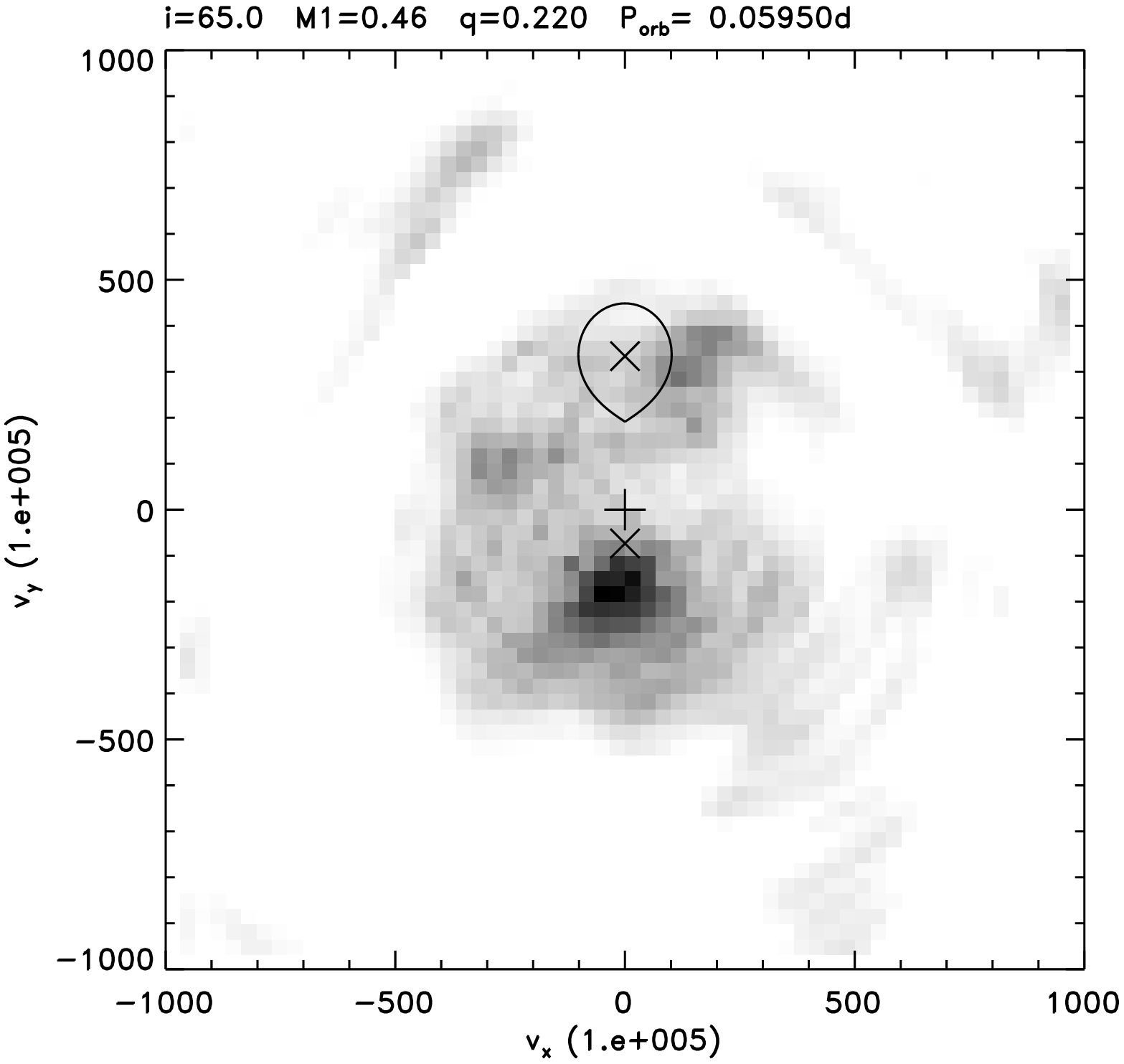}
\includegraphics{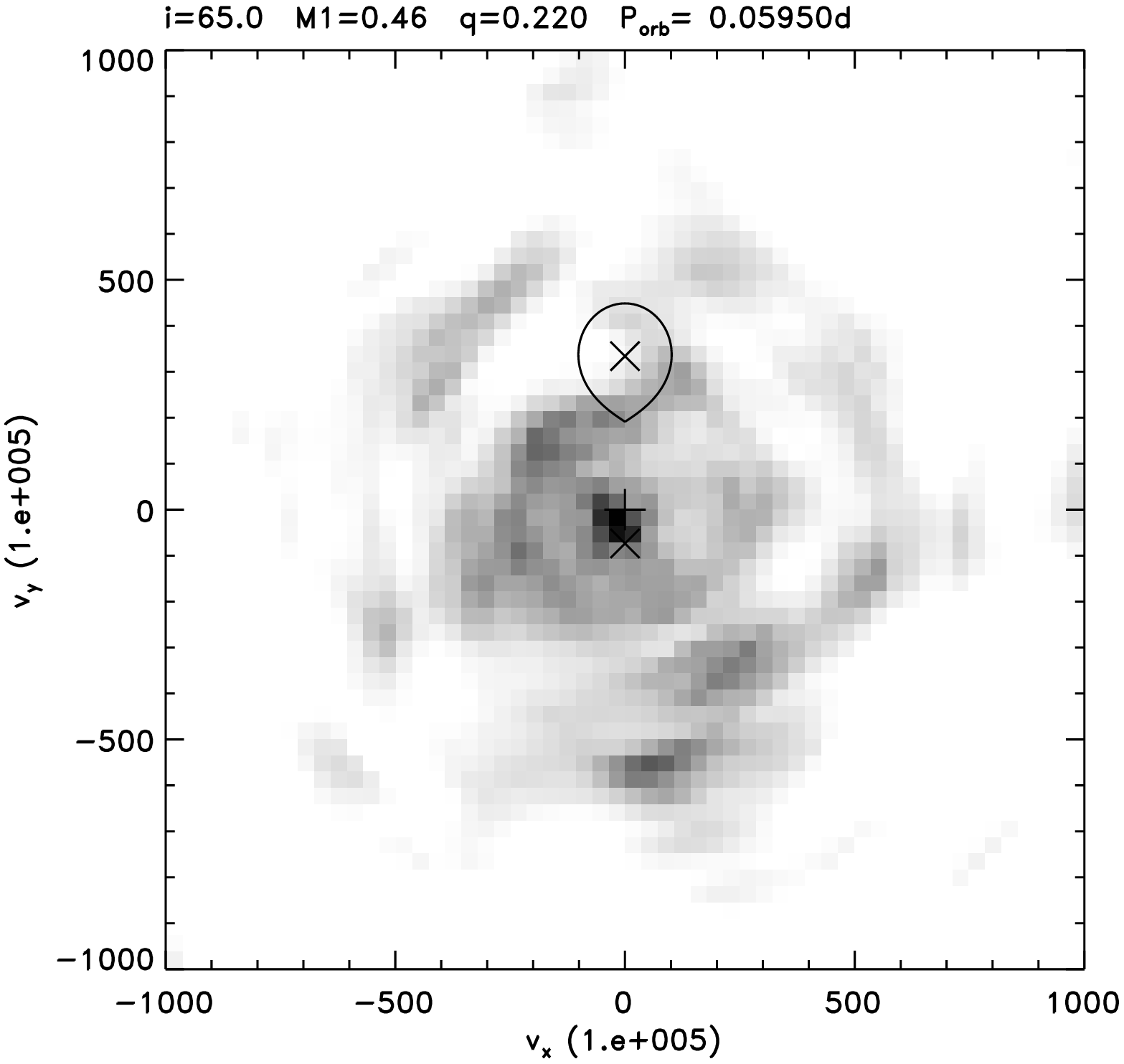}
}
\hfill
\vspace*{0.5cm}
\caption{Doppler tomograms for the He\,I$\lambda $4471 (left) and He\,II
$\lambda $4686 (right) emission lines of FS Aur. The main emission source in 
He\,I$\lambda $4471 is the first bright spot visible in H$_{\beta }$ 
(Fig.~\ref{top_bg}). The tomogram of He\,II shows an emission ring with a radius 
of about 225 km s$^{-1}$, which is centered at (V$_x \approx 0$ km s$^{-1}$, 
V$_y \approx -20$ km s$^{-1}$). In the center of this ring there is a compact bright 
spot, which is a quite bright source.}
\label{top_he}
\end{figure*}

The Maximum Entropy Doppler maps of the H$_\beta $, H$_\gamma $, 
He\,I $\lambda $4471 and He\,II $\lambda $4686 emission 
were computed using the code developed by Spruit (\cite{spruit}). The resulting 
tomograms are displayed as a gray-scale image in Figs.~\ref{top_bg} and 
\ref{top_he}. These figures also show trailed spectra in phase space and 
their corresponding reconstructed counterparts. A helpful assistance 
in interpreting Doppler maps are additional inserted plots  which mark
the positions of the white dwarf, the center of mass of the binary and the
Roche lobe of the secondary star. 
These markers are necessary to us only for facilitation of the interpretation 
of the tomograms, therefore the calculations of their positions can be done 
using our preliminary system parameters obtained in Sect.~\ref{syspar}.
Here we have used $q=0.22$, $M_{1}=0.46 M_{\sun}$ and $i=65^{\circ}$.

The Balmer Doppler maps display a roughly symmetric and very nonuniform 
distribution of the emission centered near the white dwarf.
So, on the H$_\beta $ tomogram, in addition to the symmetric distributed 
emission, at least two additional emission sources can be seen.
The first, brighter enhanced emission component is roughly centered on
(V$_x \approx -100$ km s$^{-1}$, V$_y \approx -260$ km s$^{-1}$).
The second occupies an extending area from azimuths about $300^{\circ}$ to
$60^{\circ}$ (the appropriate phase of the cross of the line-of-sight by
this bright region is about $0.85 - 1.15$). 
The first spot is well visible also on the Doppler maps of H$_\gamma $ and 
He\,I $\lambda $4471: in the case of He\,I it is a primary radiating 
source. In addition, the tomograms of H$_\gamma $ and He\,II $\lambda $4686 show 
an emission ring with radius of about 225 km s$^{-1}$, which is centered at 
(V$_x \approx 0$ km s$^{-1}$, V$_y \approx -20$ km s$^{-1}$).
In the center of this ring there is a compact bright spot, which in He\,II is a 
brightest radiating source.

The interpretation of the spot structure detected in the accretion disk is ambiguous.
Neither the first nor the second bright region can contribute to the emission 
from the bright spot on the outer edge of the accretion disk, as both areas of 
additional emission lie far from the region of interaction between the 
stream and the disk's particles. 
The brighter spot can be interpreted as due to an enhanced emission region 
located opposite to the bright spot expected by the standard model.
Earlier Mennickent (\cite{menn1994}) has shown that the bright spot region seems
``to migrate" towards the back of the disk in systems with low mass ratios.
This ``reversed bright spot" phenomenon can be explained by a gas stream
which passes above the disk and hits its back, or alternatively,
by the disk thickening in resonating locations.
We will analyze the nature of the detected structure of the accretion disk  
in the following section.

\section{Discussion}

\subsection{System's parameters}

FS Aur is a non-eclipsing binary, so information on the system's parameters
is difficult to obtain accurately. Usually for definition of the parameters of 
such systems some empirical formulae are used.
However, it should be noted that these assessments are subject to 
unknown and potentially large errors and should be adopted with
appropriate caution. In the present paper we have decided to restrict 
ourselves only to estimation of the parameters.  
To ensure some validation, we have used only stringent relations.
So, the mass-period relations for the 
secondary stars we have used also in the most facilitated form.
Thus we believe that the restrictions on the basic system parameters 
for FS Aur obtained by us are correct, if the values of input data 
are correct too.

In this connection we pay attention to the obtained estimate of the mass of the white dwarf 
which is quite small. We have estimated the primary mass to be less (and possibly 
much less) than 0.46 $M_{\sun}$, that is near to a lower limit of the observed range 
for white dwarf masses in cataclysmic variables (Webbink \cite{webbink}; 
Sion \cite{sion})\footnote{The mean mass estimate of 76 CV white dwarfs is
$M_{WD}=0.86M_{\sun}$ (Sion \cite{sion}). Webbink (\cite{webbink}) also gives 
statistically average masses for all systems below the period gap: $M_{WD}=0.61M_{\sun}$}. 
The estimate of $M_{1}$ depends strongly on the estimated $K_{1}$ value.
Let us discuss possible errors in the definition of $K_{1}$.

The definition of the semi-amplitude of the radial velocities of the emission lines,
really reflecting the orbital moving of a white dwarf, is a very complicated
problem. The contribution to a broad emission line can be introduced by
many emission areas of a binary system. For example, Balmer emission from
the secondary star forms an additional component in an emission line moving 
with semi-amplitude $K_{1}$/$q$. This component upon condition of small $q$ can 
deform the line wings. Any nonhomogeneity of the accretion disk can cause 
even greater distortion in the line profile. 

In our case the contribution of the secondary to the Doppler maps is completely
absent. At the same time, the Balmer and Helium emission is distributed 
very nonuniformly (Figs.~\ref{top_bg} and \ref{top_he}).
In general, this factor affect the accuracy of the definition of $K_{1}$, 
but we hope that it has not occurred. 
Actually, though the areas of bright inhomogeneities can be detected 
rather clearly on the Balmer tomograms, nevertheless they reach distances 
of no more than about 500 km s$^{-1}$ from the center of the tomograms.
We would like to recall, that we tested the line profiles at 650--750 
km s$^{-1}$ from its center where inhomogeneities became less noticeable.
                                                          
On the other hand it is necessary to note that TPST have found a somewhat smaller
radial velocity semi-amplitude ($K_{1}$ = 60 km s$^{-1}$). However they have
noted that the main aim of their research was the definition of the orbital period,
and their values of $K_{1}$ should not be used in dynamical solutions
of the system.
When we determined $K_{1}$ we adopted values of $\Delta$ much smaller than FWZI. 
The dependence of parameter $\sigma(K)/K$ on the Gaussian separation for H$_\beta $
and H$_\gamma $ lines is very similar (Fig.~\ref{diagram}). It can be seen that
$\sigma(K)/K$ decreases monotonically with increasing $\Delta$.
Having selected a greater value of $\Delta$ we can of course obtain a smaller value of 
$K_{1}$. However we cannot offer any convincing reasoning for increasing 
$\Delta$. It even seems undesirable to do so, as increasing $\Delta$ will 
lead us  into the farther line wings, which are subject to contortion at 
some phases (Figs.~\ref{profiles} and \ref{asymmetry}). 

Finally, it is necessary to pay attention to the emission ring, which is well 
noticeable on the H$_\gamma $ and He\,II tomograms. There is a temptation to connect it 
with the accretion disk, and the bright spot in the ring center with the white dwarf.
In this case it would become possible to independently determine $K_{1}$. 
Unfortunately, such a ring corresponds to a too large accretion disk, which 
cannot lie in the Roche lobe of the white dwarf.

Thus we believe and hope that the value of semi-amplitude of $K_{1}$ 
(and system parameters) obtained by us is correct.
Nevertheless we consider, that new, longer and better-quality observations 
are extremely necessary for a more precise definition of the system 
parameters of FS Aur.

\subsection{Accretion disk's structure}

Another important result of this work is the spot structure detected in the
accretion disk of FS Aur. Doppler tomography has shown at least two additional 
bright regions in this system. The first, brighter spot is located at phase about 0.6.
The second spot is located opposite the first and occupies an extensive area 
at phases about 0.85--1.15. 
The detected spot structure of the accretion disk is confirmed by a dependence of 
equivalent widths on orbital phase (Fig.~\ref{eqw}). The observed minima 
of EW can be due to an increase in the continuum luminosity when the enhanced emission 
region crosses the line-of-sight.

An enhancement of the emission coming from the back of the
accretion disks of some cataclysmic variables was noted by many observers
(see review by Livio \cite{livio}). Some theoretical studies indicate that
high, free flowing gas could pass over the white dwarf and hit the back side of the 
disk (Lubow \& Shu \cite{lubow:shu}; Lubow \cite{lubow}).
In this case, the azimuth angle of the region of ``secondary interaction" should be 
about $140^{\circ}$--$150^{\circ}$ for a wide range of the system's parameters,
while the distance from the accreting component will change from 0.02 to 0.18 of
the system's size, depending on the mass ratio (Lubow \cite{lubow}).
The appropriate phase of the cross of the line-of-sight by this bright 
spot must be about 0.6. This is where our observed bright spot is found! 

In addition, recent numerical hydrodynamic calculations point to a possible 
difference in the thickness of the outer edge of the accretion disk in a close 
binary system. For example, Meglicki et al. (\cite{meglicki}) identified three 
thicker regions of the disk at phases 0.2, 0.5, and 0.8. Armitage \& Livio 
(\cite{armitage:livio}) also pointed to a possible transfer of the stream matter 
above the plane of the accretion disk and an increase in the number of atoms along 
the line-of-sight relative to the average level at phases 0.1--0.2 and 
0.7--1.0. Thus, the second bright region of FS Aur can be connected with 
one of the accretion disk's thickening regions found by Meglicki et al. 
(\cite{meglicki}). The mechanism of the increase of its luminosity is 
not quite clear, but it is probably attributable to ionization by emission 
from the inner disk's region.

The nature of the ring-shaped structure visible on H$_\gamma $ and He\,II tomograms 
remains completely unintelligible. As was already noted above, this structure cannot 
be connected with the accretion disk, as its size should be so large, that it cannot 
lie in the Roche lobe of the white dwarf. Another possible site of 
origin is a nebula. However, the radial velocity of the nebula's center should coincide 
with the systemic velocity of the binary. In our case though the velocities are close, 
but still noticeably different. We again come to the conclusion, that new and more 
better quality data for the explanation of this puzzle are necessary.

\section{Summary}

In this paper we present the results of non-simultaneous time-resolved photometric 
and spectroscopic observations of the little-studied dwarf nova FS Aur in quiescence. 
We have obtained the following results:
\begin{itemize}
\renewcommand{\labelitemi}{$\bullet$}
\item 
The spectrum of FS Aur shows strong and broad emission lines of hydrogen and 
He\,I, and of weaker He\,II $\lambda4686$ and C\,III/N\,III blend, similar to 
other quiescent dwarf novae;
\item
All emission lines in the spectrum of FS Aur are single-peaked, however their 
form varies with orbital phase;
\item
Absorption lines from a late-type secondary are not detected;
\item
From the radial velocity measurements of the hydrogen lines H$_\beta$ and 
H$_\gamma$ we have determined a most probable orbital period P=0\fd059 $\pm$ 0\fd002. 
This period agrees well with the 0\fd0595 $\pm$ 0\fd0001 
estimate by TPST. On the other hand, the period of 
photometric modulations is longer than the spectroscopic period and can be 
estimated as 3 hours. Longer time coverage during a single night is 
needed to resolve this problem;
\item
Using the semi-amplitude of the radial velocities, obtained from measurements
of hydrogen and helium lines, and some empirical and theoretical relations 
we limited the basic parameters of the system: the mass ratio $q \geq 0.22$, 
the primary mass 
$M_{1}=0.34-0.46 M_{\sun}$, the secondary mass $M_{2} \leq 0.1M_{\sun}$,
and the inclination angle $i=51^{\circ }-65^{\circ}$;
\item
Doppler tomography has shown at least two bright regions in the accretion disk 
of FS Aur. The first, brighter spot is located at a phase of about 0.6.
The second spot is located opposite the first one and occupies an extensive area 
at phases about $0.85 - 1.15$. 
\end{itemize}

\begin{acknowledgements}
I am grateful to Oksana van den Berg for contributing to a improved first version of 
this paper. I would like to thank the firm VEM (Izhevsk, Russia) and Konstantin Ishmuratov 
personally for the financial support rendered to me in the preparation of this paper.
Thanks also to Alexander Khlebov for the computer and technical support. I  
acknowledge an anonymous referee for detailed reading of the manuscript, 
and useful suggestions concerning the final version.
\end{acknowledgements}


\begin{thebibliography}{}

\bibitem[1991]{andronov} Andronov I.I., 1991, IBVS, No. 3614
\bibitem[1996]{armitage:livio} Armitage P.J., Livio M., 1996, ApJ 470, 1024
\bibitem[1990]{cail:patt} Caillault J.P., Patterson J., 1990, AJ 100, 825
\bibitem[1961]{hamada:salpeter} Hamada T., Salpeter E.E., 1961, ApJ 134, 683
\bibitem[1949]{hoff} Hoffmeister C., 1949, Veroff. Sternw. Sonneberg 1, 3
\bibitem[1986]{horne} Horne K., 1986, PASP 98, 609
\bibitem[1988]{how:szkody} Howell S.B., Szkody P., 1988, PASP 100, 224
\bibitem[1993]{livio} Livio M., 1993, In: ``Accretion Disks in Compact Stellar Systems",
                      Ed. J.C. Wheeler (Singapore: World Sci. Publ.), p.243
\bibitem[1976]{lubow:shu} Lubow S.H., Shu F.H., 1976, ApJ 207, L53
\bibitem[1989]{lubow} Lubow S.H., 1989, ApJ 340, 1064
\bibitem[1988]{marsh:horne} Marsh T.R., Horne K., 1988, MNRAS 235, 269
\bibitem[2001]{marsh2001} Marsh T.R., 2001, in ``Proceedings of the Astro-Tomography
Workshop", eds H. Boffin, D. Steeghs, in press
\bibitem[1993]{meglicki} Meglicki Z., Wickramasinghe D., Bicknell G.V., 1993,
               MNRAS 264, 691
\bibitem[1994]{menn1994} Mennickent R.E., 1994, A\&A 285, 979
\bibitem[1996]{misselt} Misselt K.A., 1996, PASP 108, 146
\bibitem[1984]{patterson} Patterson J., 1984, ApJS 54, 443
\bibitem[1980]{sch:young} Schneider D.P., Young P., 1980, ApJ 238, 946
\bibitem[1983]{shafter1} Shafter A.W., 1983, ApJ 267, 222
\bibitem[1984]{shafter2} Shafter A.W., Szkody P., 1984, ApJ 276, 305
\bibitem[1986]{shafter3} Shafter A.W., Szkody P., Thorstensen J.R.,
                         1986, ApJ 308, 765
\bibitem[1999]{sion} Sion E.M., 1999, PASP, 111, 532
\bibitem[1998]{smith:dhillon} Smith D.A., Dhillon V.S., 1998, MNRAS 301, 767
\bibitem[1998]{spruit} Spruit H.C., 1998, astro-ph/9806141
\bibitem[1996]{TPST} Thorstensen J.R., Patterson J.O., Shambrook A., Thomas G.,
1996, PASP 108, 73 (TPST)
\bibitem[1995]{warner95} Warner B., 1995, Cataclysmic Variable Stars.
             (Cambridge Astrophysics Ser. 28; Cambridge: Cambridge Univ. Press)
\bibitem[1990]{webbink} Webbink R.F., 1990, in: ``Accretion Powered Compact Binaries",
             Ed. C.W. Mauche (Cambridge: Cambridge Univ. Press), p.177
\bibitem[1983]{williams} Williams G., 1983, ApJS 53, 523


\end{thebibliography}
\end{document}